\documentclass[11pt,twocolumn]{article}
\usepackage{longtable}
\usepackage{amsmath}
\usepackage{amssymb}
\usepackage{graphicx}
\usepackage{bm}
\usepackage{footmisc}
\usepackage{afterpage}
\usepackage{float}
\usepackage{lscape}
\usepackage{longtable}
\usepackage{float}
\usepackage{slashed}
\usepackage{morefloats}
\usepackage{afterpage}
\usepackage{abstract}
\usepackage{appendix}
\usepackage{mathtools, cuted}
\usepackage{comment}
\usepackage{braket}
\usepackage{authblk}
\usepackage{titlesec}

\newcommand*{\no}{\noindent}
\newcommand*{\bea}{\begin{eqnarray}}
\newcommand*{\eea}{\end{eqnarray}}
\newcommand*{\be}{\begin{equation}}
\newcommand*{\ee}{\end{equation}}

\newcommand*{\pref}[1]{(\ref{#1})}

\newcommand*{\mn}{{\mu\nu}}

\newcommand*{\nn}{\nonumber}

\newcommand*{\tr}{\mathrm{tr}}

\newcommand{\bma}{\begin{pmatrix}}
	\newcommand{\ema}{\end{pmatrix}}

\titlespacing*{\section}{0cm}{10pt}{10pt}
\usepackage[square,comma,sort&compress,numbers]{natbib}

\title{A composite massless vector boson}

\author{Vincenzo Afferrante}

\author{Axel Maas}

\author{Pascal T{\"o}rek}

\affil{Institute of Physics, NAWI Graz, University of Graz, Universit\"atsplatz 5, A-8010 Graz, Austria}

\begin{document}

	\maketitle
\begin{strip}
	\begin{abstract}
	
    In a non-perturbative gauge-invariant formulation of grand-unified theories all low energy vector states need to be composite with respect to the high-scale gauge group, including the photon. We investigate this by using lattice methods to spectroscopically analyze the vector channel in a toy grand-unified theory, an SU(2) adjoint Higgs model. Our results support indeed the existence of a massless composite vector particle.

\end{abstract}
\end{strip}

\maketitle

\section{Introduction}

Observable particles need to be described by manifestly gauge-invariant operators. Beyond perturbation theory, BRST symmetry is insufficient to ensure this in non-Abelian gauge theories. Instead composite operators are needed, irrespective of the actual value of any coupling constants \cite{Frohlich:1980gj,Frohlich:1981yi,Maas:2017wzi}. Thus, in electroweak physics an identification of the observed particles and the elementary, gauge-dependent degrees of freedom of the Lagrangian is not directly possible. However, due to a combination of the Brout-Englert-Higgs (BEH) effect together with the Fr\"ohlich-Morchio-Strocchi (FMS) mechanism \cite{Frohlich:1980gj,Frohlich:1981yi} this happens effectively, up to corrections suppressed by powers of the Higgs vacuum expectation value. This has been confirmed in lattice calculations \cite{Maas:2012tj,Maas:2013aia,Maas:2017wzi}, including subleading contributions \cite{Maas:2018ska,Sondenheimer:unpublished,Maas:2019dwd}, and has potentially experimentally observable consequences \cite{Maas:2018ska,Egger:2017tkd,Fernbach:unpublished,Maas:2019dwd}. For a review see \cite{Maas:2017wzi}.

However, this can potentially change for theories with a different structure than the standard model, in particular in scenarios for new physics \cite{Maas:2015gma,Maas:2017xzh,Sondenheimer:2019idq}. Especially, the physical observable spectrum of particles can differ qualitatively from the one of the elementary particles, and thus from those in perturbation theory. This has also been supported in lattice calculations \cite{Maas:2016ngo,Maas:2018xxu}. Though this does not invalidate new physics scenarios as such, it does require to take manifest gauge invariance in their construction into account, by augmenting perturbation theory with the FMS mechanism \cite{Maas:2017wzi}. This yielded in all cases tested so far on the lattice \cite{Maas:2016ngo,Maas:2018xxu} correct predictions \cite{Maas:2016ngo,Maas:2017xzh} even when conventional perturbation theory did not.

In the context of grand-unified theories (GUTs) \cite{Georgi:1974sy,Langacker:1980js}, this program faces a particular challenge when it comes to model-building \cite{Maas:2017xzh,Sondenheimer:2019idq}. In GUTs all low-energy interactions are created from a single non-Abelian gauge group, including QED\footnote{Gauge-invariance is already non-trivial in QED, where a Dirac string is needed to make the photon gauge-invariant with respect to the electromagnetic gauge group \cite{Maas:2017wzi,Lavelle:1995ty}. This has been confirmed in lattice simulations, see e.\ g.\ \cite{Woloshyn:2017rhe,Lewis:2018srt}. This can be included straight-forwardly into the FMS description of the electroweak sector of the standard model \cite{Shrock:1986av,Maas:2017wzi}, which is again confirmed by lattice investigations \cite{Shrock:1985un,Shrock:1985ur}.}. This requires the presence of a massless, uncharged vector particle, which is composite with respect to the GUT gauge group, to play the role of the low-energy photon\footnote{Note that composite massless photons appear also in non-GUT contexts, e.\ g.\ as bound states of new fermions \cite{Terazawa:1976xx,Terazawa:2015bsa}.}. FMS-augmented perturbation theory indeed predicts that such states can arise when adjoint Higgs fields are present \cite{Maas:2017xzh}. And, in fact, early exploratory lattice investigations seem to support the presence of such a composite massless vector particle \cite{Lee:1985yi}. Our aim is to substantiate these results. In addition, also the massive vector states are predicted \cite{Maas:2017xzh} to differ from those of perturbation theory. Thus, we also test this.

To this end, we will simulate the simplest theory which is expected to show this behavior, SU(2) Yang-Mills theory with a single Higgs in the adjoint representation. We will discuss this theory and the relevant predictions, both of perturbation theory and the FMS mechanism, in Section \ref{sec:continuum}. Our lattice implementation will be given in Section \ref{sec:setup}, with some details relegated to appendix \ref{Apppendix:therm}. In particular, we found that early investigations of the phase diagram of this theory \cite{Brower:1982yn,Gupta:1983zv,Drouffe:1984hb,Baier:1986ni,Baier:1988sc} likely underestimated systematic effects due to the finite volume and length of Monte Carlo trajectories, similar to what has happened in the fundamental case \cite{Bonati:2009pf}. These effects are quite severe, and thus also in our case we cannot yet offer a full systematic analysis in terms of discretization artifacts, though volume effects will be investigated in great detail.

Since perturbation theory uses gauge-fixed calculations, we need to replicate this on the lattice to provide the corresponding results for comparison. For this, we use the minimal 't Hooft-Landau gauge \cite{Maas:2016ngo,Maas:2017wzi}. This also allows us to determine the running gauge coupling in the miniMOM scheme \cite{vonSmekal:2009ae}, and to compare gauge-fixed correlation functions to their perturbative predictions. By this we verify that we indeed work at weak coupling. This is also a necessary step to obtain the FMS predictions \cite{Maas:2017xzh,Maas:2017wzi}. This is discussed in Section \ref{sec:gauge_fixed}.

Finally, the central result is the spectroscopical analysis of the vector channel in Section \ref{sec:results}, which is obtained with the methods described in Section \ref{sec:Gauge_invariant}. The spectrum is found to be compatible with the results from the FMS mechanism \cite{Maas:2017xzh}. Especially, we find the massless composite vector state, which would act as the photon in a GUT scenario. We do not find evidence for further massive states. These findings are summarized and put into perspective in Section \ref{sec:conclusion}. Some preliminary results can be found in \cite{Afferrante:2019vsr}.


\section{Continuum $\mathrm{SU}(2)$ theory coupled to an adjoint scalar}\label{sec:continuum}

The theory we investigate is described by the Lagrangian
\begin{equation}
\mathcal{L} =  -\dfrac{1}{4}W^a_{\mu \nu} W^{a \mu \nu} + \tr \big[(D_\mu \Phi)^\dag(D^\mu \Phi)\big] -V(\Phi) \;.\nn
\end{equation} 
The scalar field can be expanded as  $\Phi(x)  = \Phi^a(x) T^a$, where $T^a$ are the generators of the Lie algebra of the group. The components $\Phi^a$ form a three dimensional real-valued vector. The scalar field transforms under a gauge transformation $G$ as $\Phi(x) \rightarrow G(x) \Phi(x) G(x)^\dag $. The covariant derivative acts as $D^\mu \Phi = \partial^\mu \Phi + i g [W^\mu, \Phi]\,$. The gauge fields $W_\mu=W_\mu^a T^a$ and their field strength tensor $W_\mn=W_\mn^a T^a$ are the usual ones of Yang-Mills theory.

The potential is taken to be the most general one renormalizable by power-counting and conserving the $Z_2$ transformation $\Phi\to-\Phi$,
\begin{equation}
V(\Phi) =  -\mu^2\, \tr\big[ \Phi\big]^2 + \dfrac{\overline{\lambda}}{2}\, \tr \big[\Phi^2\big]^2 \, + \tilde{\lambda} \,\tr \big[\Phi^4\big]\; . \label{pot}
\end{equation} 
\no However, in the case of the gauge group SU(2) $\tr \Phi ^2$ is the only nontrivial invariant Casimir, and we can therefore combine the last two terms into one with a single coupling constant $\lambda$. In addition, because the field is in the not-faithful adjoint representation of the pseudo-real SU(2), the $Z_2$ symmetry is not an independent field transformation when the theory is gauged. Hence, there is no global (custodial) symmetry, and there are no global quantum numbers in this theory except for spin and parity.

\subsection{Gauge-variant description in a fixed gauge}\label{sub:gauge_dep_obs}

To test the FMS mechanism and compare to usual perturbative treatments it is necessary to consider the gauge-fixed theory. Since our interest is the BEH domain, only this case will be considered. For the present theory, the (only) breaking pattern is SU(2) $\rightarrow$ U(1) \cite{Bohm:2001yx}, i.e., an unbroken U(1) subgroup is left.

It is then possible to choose a suitable gauge, here minimal 't Hooft-Landau gauge \cite{Maas:2017wzi}, where the scalar field can be split into a constant and a fluctuating part, i.e., 
\begin{equation}
\Phi(x) = \braket{\Phi} + \phi(x) \equiv w\,\Phi_0 + \phi(x) \;.\label{split}
\end{equation}
\no $\Phi_0$ is the direction of the vacuum expectation value obeying $\Phi_0^a \Phi_0^a = 1$, and $w$ is its magnitude.  $\Phi_0$ can always be chosen inside the Cartan \cite{Bohm:2001yx}. Gauge transformations in the unbroken U(1) subgroup leave $\Phi_0$ invariant. The field $\phi=\phi^aT^a$ is the fluctuation field.

Inserting the split \pref{split} into the Lagrangian yields the tree-level mass matrix
\begin{equation*}
(M^2_A)^{a b} = - 2 (gw)^2 \tr \left([T^a,\Phi_0][T^b,\Phi_0]\right)\;,
\end{equation*} 
for the gauge bosons. This leads to a massless gauge field for the unbroken U(1) subgroup. The masses of the two SU(2) coset gauge bosons are $m_A= g w$. In addition, one degree of freedom of the scalar Higgs field remains with mass $m_H = \sqrt{\lambda} w$.

\subsection{Gauge-invariant spectrum}\label{sub:gauge_invariant_obs}

As discussed in the introduction, the observable spectrum needs to be manifestly and non-perturbatively gauge-invariant \cite{Frohlich:1980gj,Frohlich:1981yi}. For the present theory this spectrum has been predicted in \cite{Maas:2017xzh} for the $0^+$ and $1^-$ channels, implying the presence of non-scattering states in both. For completeness, we will rehearse here the predictions of \cite{Maas:2017xzh} for these two channels.

Consider first the $0^+$ channel. The simplest composite gauge-invariant operator with these quantum numbers is
\begin{equation}
O_{0^+}(x) = \tr \big[\Phi^2\big](x)\nn\;.
\end{equation}
\no To obtain the leading-order prediction for the associated mass spectrum for this operator, FMS-augmented perturbation theory requires to expand this operator in the vacuum expectation value $w$ to leading non-constant order \cite{Maas:2017wzi}, yielding
\begin{equation}
O_{0^+}(x) = \dfrac{w^2}{2} + w\,H(x)+{\cal O}(w^0)\;,\label{higgs}
\end{equation}
\no with the Higgs field $H(x)=2\tr(\Phi_0\phi(x))$. Thus, at this order, the operator is, up to an irrelevant constant, identical to the Higgs. States created by this operator should thus have the same mass spectrum as the elementary Higgs. Especially, at tree-level the scalar singlet should have the mass of the Higgs at tree-level, i.e., $m_H$.

The situation is somewhat more involved for the vector channel $1^-$. Because of the special features of the present theory, the simplest operator, generalized from the fundamental case \cite{Frohlich:1980gj,Frohlich:1981yi}, is \cite{Maas:2017xzh}
\begin{equation}
O^\mu_{1^-} = \dfrac{\partial_\nu}{\partial^2} \tr \big[\Phi F^{\mu \nu}\big] \;.\label{eq:cont_operator}
\end{equation}
Performing the same expansion yields \cite{Maas:2017xzh}
\begin{equation}
O^\mu_{1^-} = - w \,\tr \big[\Phi_0 \overline{W}_\perp^\mu\big](x) + \mathcal{O}(w^0) \;,\label{fmsv}
\end{equation}
with
\be
\overline{W}_\perp^\mu=W_\perp^\mu+g\dfrac{\partial_\nu}{\partial^2}[W^\mu,W^\nu]\;,\label{fperp}
\ee
the field-strength tensor with one index transversely contracted and 
\be
W_\perp^\mu = \left(\delta^\mu_\nu - \frac{\partial^\mu \partial_\nu}{\partial^2}\right) W^\nu\;,\nn
\ee
the transverse part of the gauge field.

At tree-level \pref{fmsv} reduces to
\begin{equation}
O^\mu_{1^-} = - w\, \tr \big[\Phi_0 W_\perp^\mu\big](x) + \mathcal{O}(w^0,g^0,\lambda^0) \;. 
\end{equation}
The trace with $\Phi_0$ projects precisely to the transverse gauge boson of the unbroken U(1) subgroup. Thus, the state created by this operator should contain a massless pole. Hence, this predicts \cite{Maas:2017xzh} a massless, composite vector boson. This gauge-invariant state could potentially play the role of an effective low-energy photon in a GUT setup.

At leading order in $w$, but next-to-leading order in $g$, this changes. While the first term in \pref{fperp} will give rise only to a scattering threshold, this is no longer obvious for the second term. A detailed analysis in a constituent-like evaluation \cite{Maas:2017xzh} yields that a second pole at $2m_A$ could arise, and thus a second, massive vector particle. Of course, such a particle, like the scalar, will not be stable against decay into the massless vectors, but the level can still show up in the spectrum as a resonance, if it is present and decays weakly enough.

Unfortunately, it turns out that the scalar is far too noisy to obtain reliable results with about five million core hours of computing time available to us in this project. The reason is that it has vacuum quantum numbers, and thus suffers from the presence of disconnected contributions. This substantially enlarges the noise. Though we saw a signal in the lattice simulations presented here at short times, the signal drowned to quickly in noise to determine spectral information. We estimate that at least an order of magnitude more statistics, and probably further improved operators, will be necessary for a result of similar quality as in the vector channel.

Thus, we will concentrate here only on the predictions in the vector channel. In principle, there could also be non-scattering states in other channels. But because of the lack of elementary particles with other spin-parity quantum numbers no one-to-one mapping in the sense of the FMS mechanism, e.\ g\ as in \pref{higgs} for the $0^+$ channel and the Higgs, is possible. They would therefore be non-trivial bound states, and could be searched for along the lines of \cite{Wurtz:2013ova,Maas:2014pba} in the fundamental case. Based on the experience with these cases, this will likely require substantially more statistics than even for the $0^+$, and we will leave these others channels therefore to future investigations.


\section{Setup}\label{sec:setup}

\subsection{Lattice setup and parameters}

The lattice action can be obtained by discretization of the action as \cite{Montvay:1994cy}
\begin{align*}
S[\Phi&,U] = S_W[U]\\
&+ \sum_x \Big(2\,\tr\big[\Phi(x)\Phi(x)\big] \\
&+ \lambda \big(2\,\tr \big[\Phi(x) \Phi(x)\big] - 1\big)^2  \\
&- 2\, \kappa \sum_{\mu=1}^4 \tr\big[\Phi(x) \,U_\mu(x)\, \Phi(x + \hat \mu) \,U_\mu(x)^{\dagger} \big]\Big)\;,
\end{align*} 
with $S_W$ the standard Wilson action and $U_\mu(x)$ are the usual links. The action can be rewritten in component form:
\begin{align*}
S[\Phi&,U] =  S_W[U] \\
&+\sum_{x}\Bigg[\sum_{a=1}^3\Big( \Phi^a(x) \Phi^a(x) \\
&\phantom{\sum\sum }+ \lambda\,\big(\Phi^a(x) \Phi^a(x) -1 \big)^2\Big)\\
&- 2\,\kappa \sum_{\mu=1}^4\sum_{a,b=1}^3 \Phi^a(x)\,V_\mu^{a b} (x)\,\Phi^b(x + \hat \mu) \Bigg]\;,
\end{align*}
with 
\begin{equation}
V_\mu^{a b}(x) = \tr\big[T^a\,U_\mu(x)\,T^b\,U_\mu(x)^\dagger\big]\;,\nonumber
\end{equation} 
which are the links in the adjoint representation. In fact, the latter form of the action has been used for our simulations.

Lattice of sizes $L^4=8^4$, $12^4$, $16^4$, $20^4$, $24^4$, and $32^4$ have been used. For the simulation, a multi-hit Metropolis Monte-Carlo algorithm has proven to be effective for the purpose of generating the configurations, like in \cite{Maas:2016ngo,Maas:2017xzh}, see also appendix \ref{Apppendix:therm}. For every update of the scalar field five updates of the gauge field have been employed, and five hits have been used for every update. This created a new configuration.

\subsection{Phase diagram and simulation points}\label{sub:pd}

We have scanned, similarly to \cite{Maas:2016ngo,Maas:2018xxu}, a wide range of lattice parameters within the $(\beta,\kappa, \lambda)$ volume. However, we encountered severe critical slowing down. This is discussed in detail in appendix \ref{Apppendix:therm}. Especially, we found that with better thermalization properties the results on the phase diagram from exploratory investigations \cite{Brower:1982yn,Gupta:1983zv,Drouffe:1984hb,Baier:1986ni,Baier:1988sc} changed, and especially the phase transition shifted to larger values of $\kappa$ for larger volumes. The reason for this is likely the presence of the massless gauge-invariant vector particle, and thus slow decorrelation and large finite-volume effects.

However, these results, together with our own, suggest a transition from a QCD-like phase to a BEH phase at any fixed values of $\beta$ and $\lambda$ when increasing $\kappa$ sufficiently. Based on the scan, and since we do aim at a proof-of-principle, we thus decided to fix $\beta=4$ and $\lambda=1$, and perform a scan in $\kappa$ from $\kappa=1/8$, i.e., a tree-level massless scalar, to $\kappa=2$. As will be seen, we find a transition at about $\kappa\approx 0.5$ between both phases, and thus concentrate primarily on the range $\kappa\in[0.5,0.7]$. For our spectroscopical analysis we take as special cases $\kappa=0.55$ and $\kappa=0.65$.

\begin{figure*}[h]
	\centering
	\includegraphics[width=0.5\textwidth]{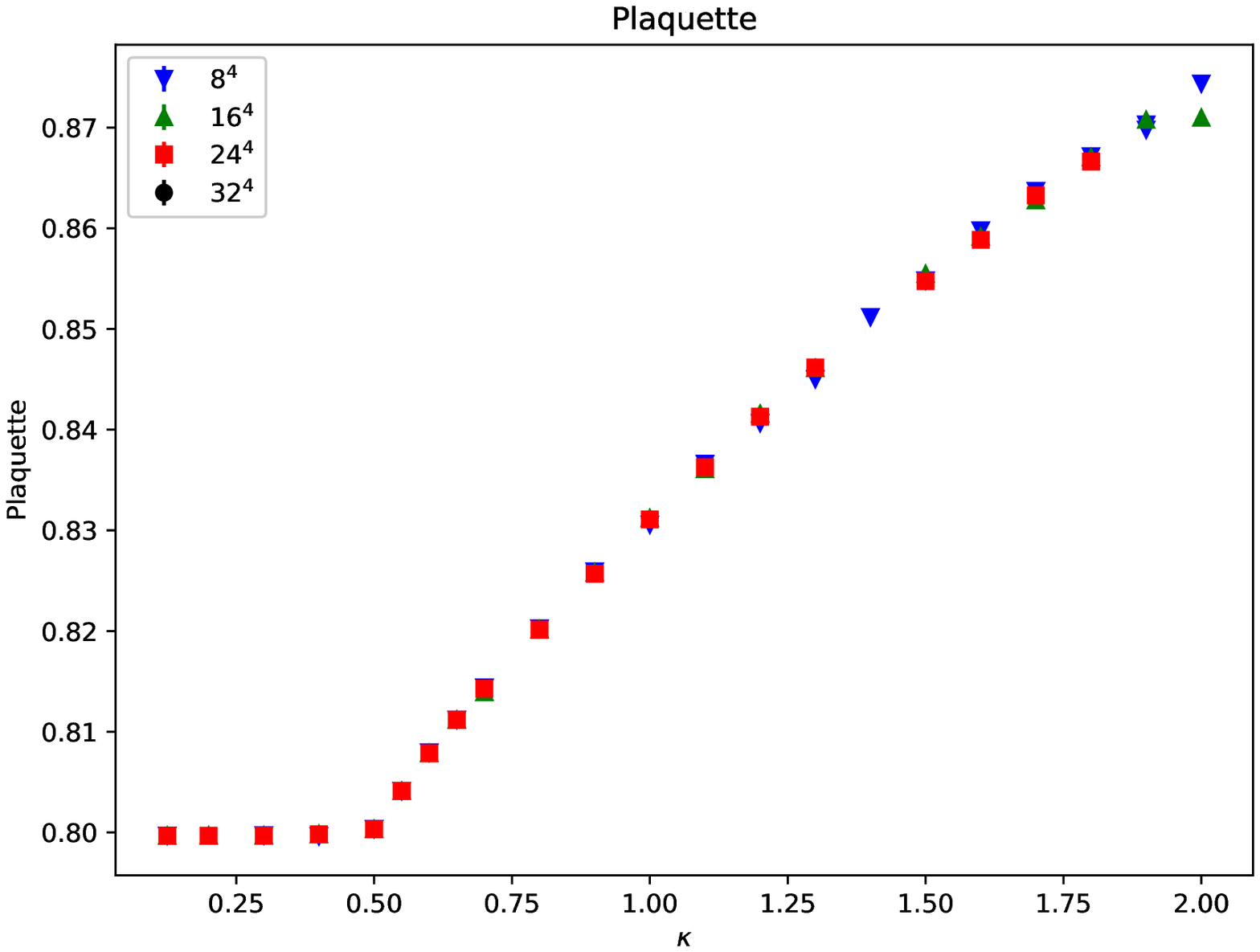}\includegraphics[width=0.5\textwidth]{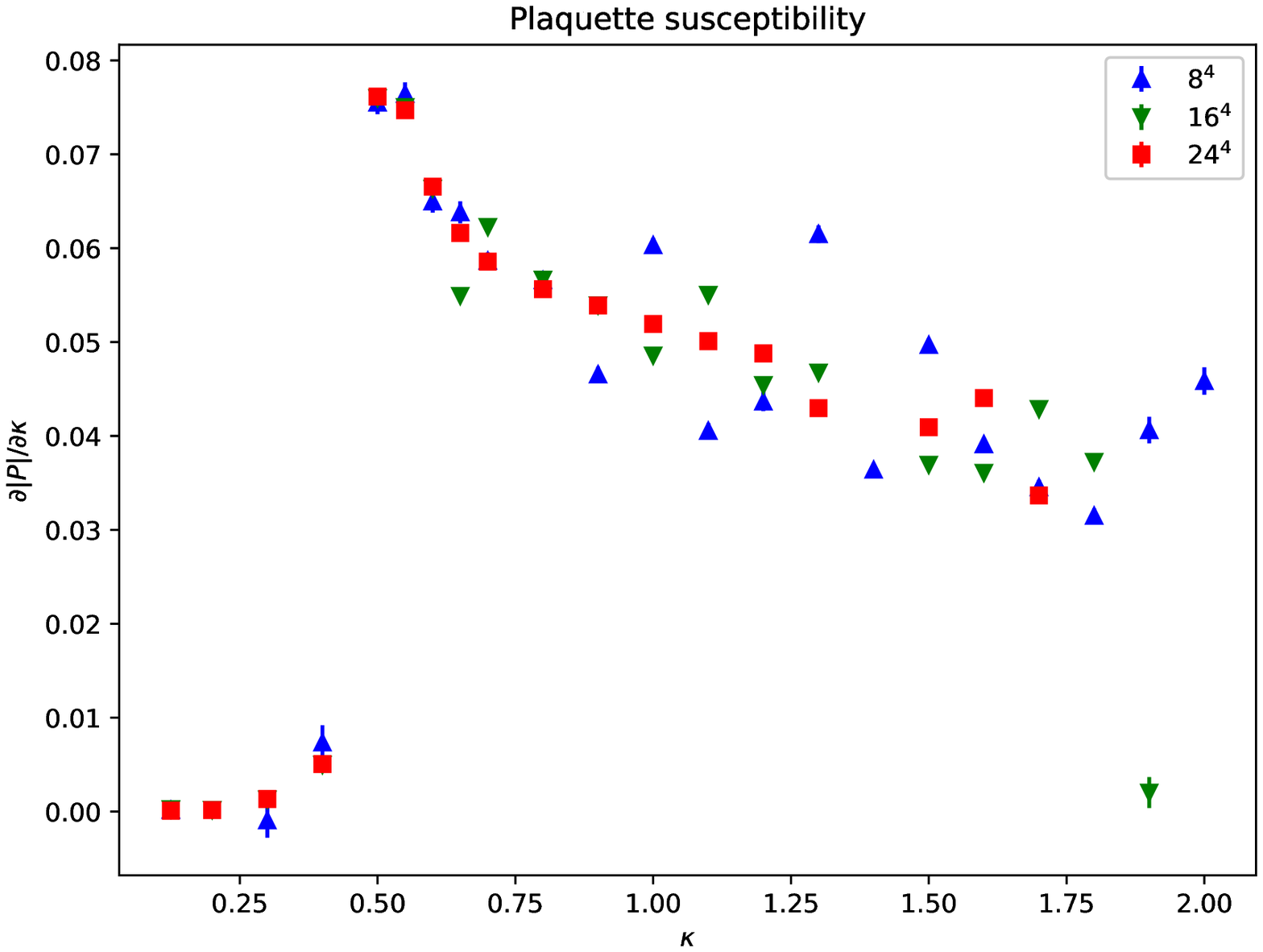}
	\caption{The plaquette as a function of $\kappa$ (left panel) for various volumes, as well as its derivative with respect to $\kappa$ (right panel). The scatter of the susceptibility at large values of $\kappa$ is an artifact of the critical slowing down discussed in appendix \ref{Apppendix:therm}.}
	\label{fig:ph_diag}
\end{figure*}

For the purpose of thermalization, we drop 1000 configurations, and drop 50 configurations for decorrelation between measurements. This is sufficient to decorrelate the plaquette, but for $\kappa\gtrsim 0.7$ not sufficient for a full decorrelation of other observables. This is discussed in more detail in appendix \ref{Apppendix:therm}. As is shown in figure \ref{fig:ph_diag}, the plaquette shows a behavior characteristic for a rapid transition around $\kappa\approx 0.5$. However, the susceptibility suggests either a cross-over or at least a very small critical region for a phase transition, due to the absence of volume scaling. Although being close to an actual second-order transition point\footnote{Even if no genuine second-order phase transition exists, we expect \cite{Hasenfratz:1986za} that low-energy observables are sufficiently reliable, just as is the case with the standard model Higgs sector \cite{Maas:2017wzi}.}, if it exists, would be preferable for a better approach to large correlation lengths, for the purpose at hand it will be sufficient to have sufficiently large correlation lengths. As will be seen, our choice of large-statistics simulation points, $\kappa=0.55$ and $\kappa=0.65$, indeed provide suitable conditions.

In total, we have simulated then 12 lattice setups in detail: For each $\kappa=0.55$ and $\kappa=0.65$ we used six lattice volumes, $8^4$, $12^4$, $16^4$, $20^4$, $24^4$, and $32^4$. For the gauge-invariant states, we used $(1-4)\times 10^5$ configurations for the smaller volumes, $8^4$ and $12^4$, and $(1-3) \times 10^4$ for the larger volumes, while for the gauge-fixed calculations an order of magnitude less configurations was used. This was necessary to compensate for the substantially increased computing time for gauge fixing, which increases with volume by one to two orders of magnitude in comparison with the generation of not gauge-fixed configurations. However, as the elementary gauge-fixed observables contain less field operators than the composite gauge-invariant ones, a similar level of statistical accuracy was nonetheless achieved, as it is expected from results on gauge dependant observables in Yang Mills theories \cite{Maas:2011se,Boucaud:2011ug}.

\section{Gauge-fixed observables}
\label{sec:gauge_fixed}\label{sec:gf_results}

As Section \ref{sub:gauge_invariant_obs} shows, testing the FMS mechanism requires information from the gauge-dependent spectrum. We therefore fix a subset of the configurations to minimal Landau-'t Hooft gauge. This is done like in \cite{Maas:2016ngo,Maas:2018xxu}, by first fixing minimal Landau gauge, and then performing a global gauge transformation to satisfy the 't Hooft gauge condition by rotating the expectation value of the Higgs field into the Cartan. In a finite volume this is always possible, even in a QCD phase, where the vacuum expectation value in any gauge vanishes in the infinite-volume limit.

Once fixed, we calculate separately the gauge boson propagators in the Cartan direction and in the remainder direction, as in \cite{Maas:2018xxu}. Furthermore, we calculate the ghost propagator to determine the running gauge coupling in the miniMOM scheme \cite{vonSmekal:2009ae}, again as in \cite{Maas:2018xxu}. This allows us to verify that we are indeed in a weak coupling regime. Finally, we also investigated the scalar boson propagator to confirm the existence of the Goldstone boson, as in \cite{Maas:2018xxu}.

\begin{figure*}
	\centering
	\includegraphics[width=0.5\textwidth]{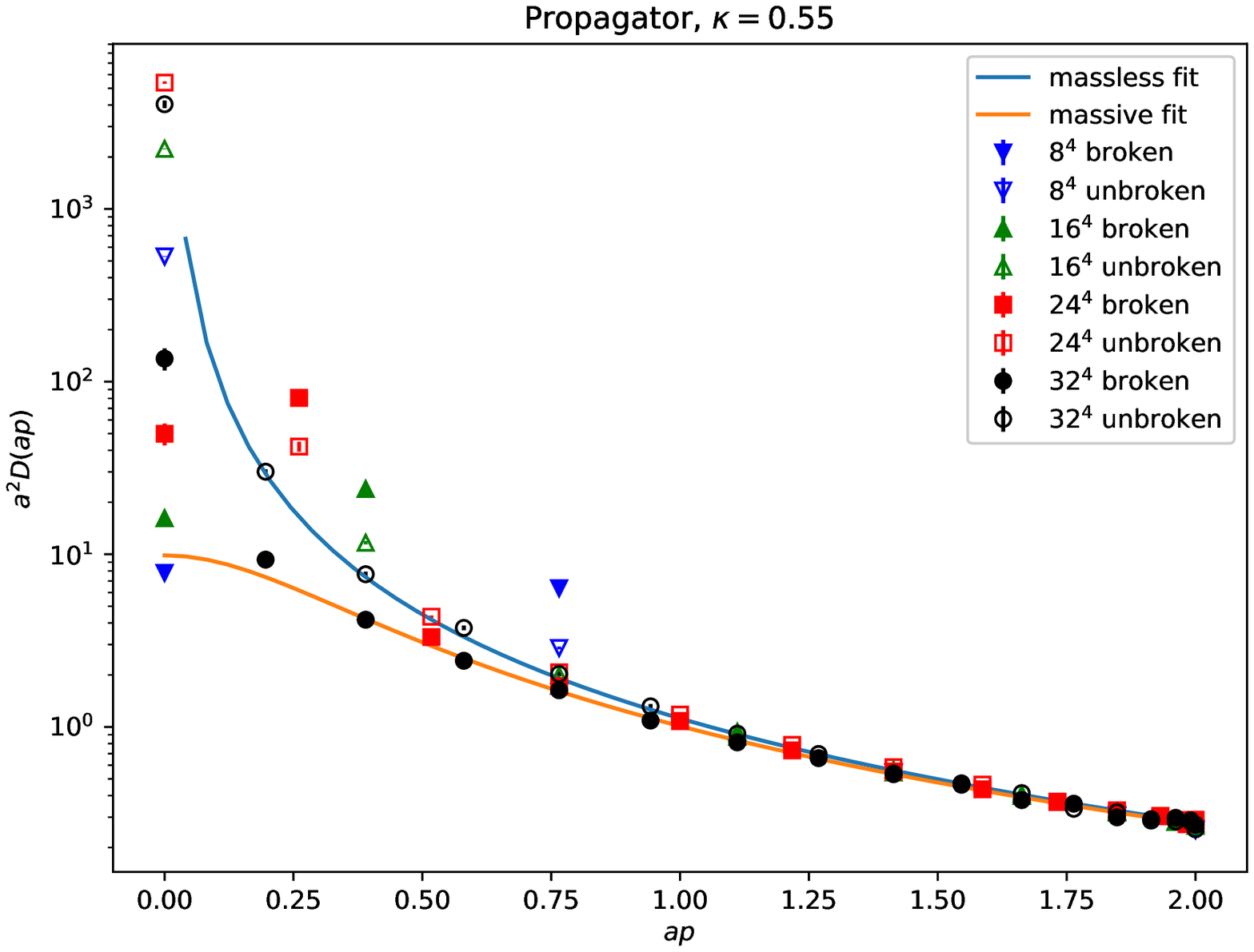}\includegraphics[width=0.5\textwidth]{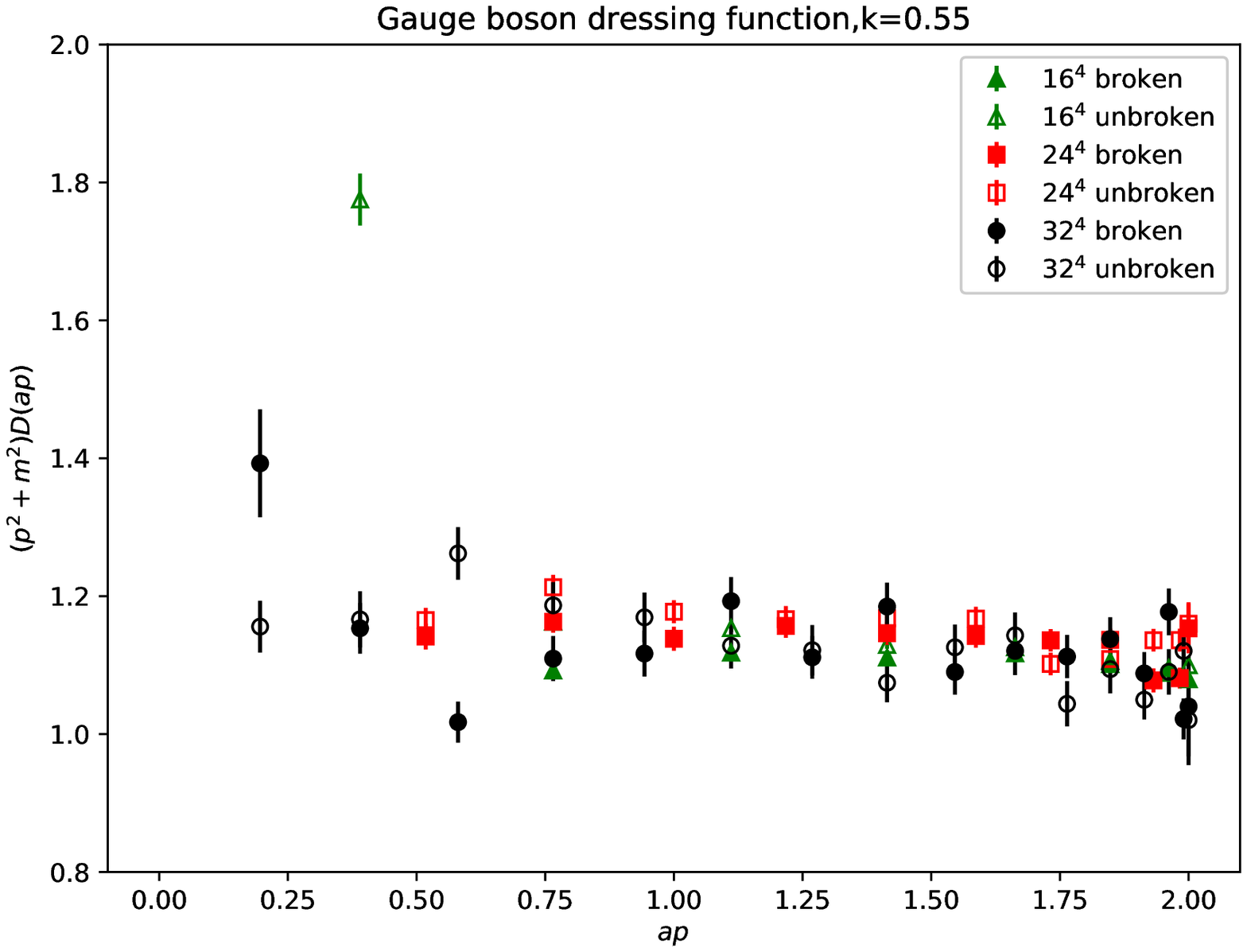}\\
	\includegraphics[width=0.5\textwidth]{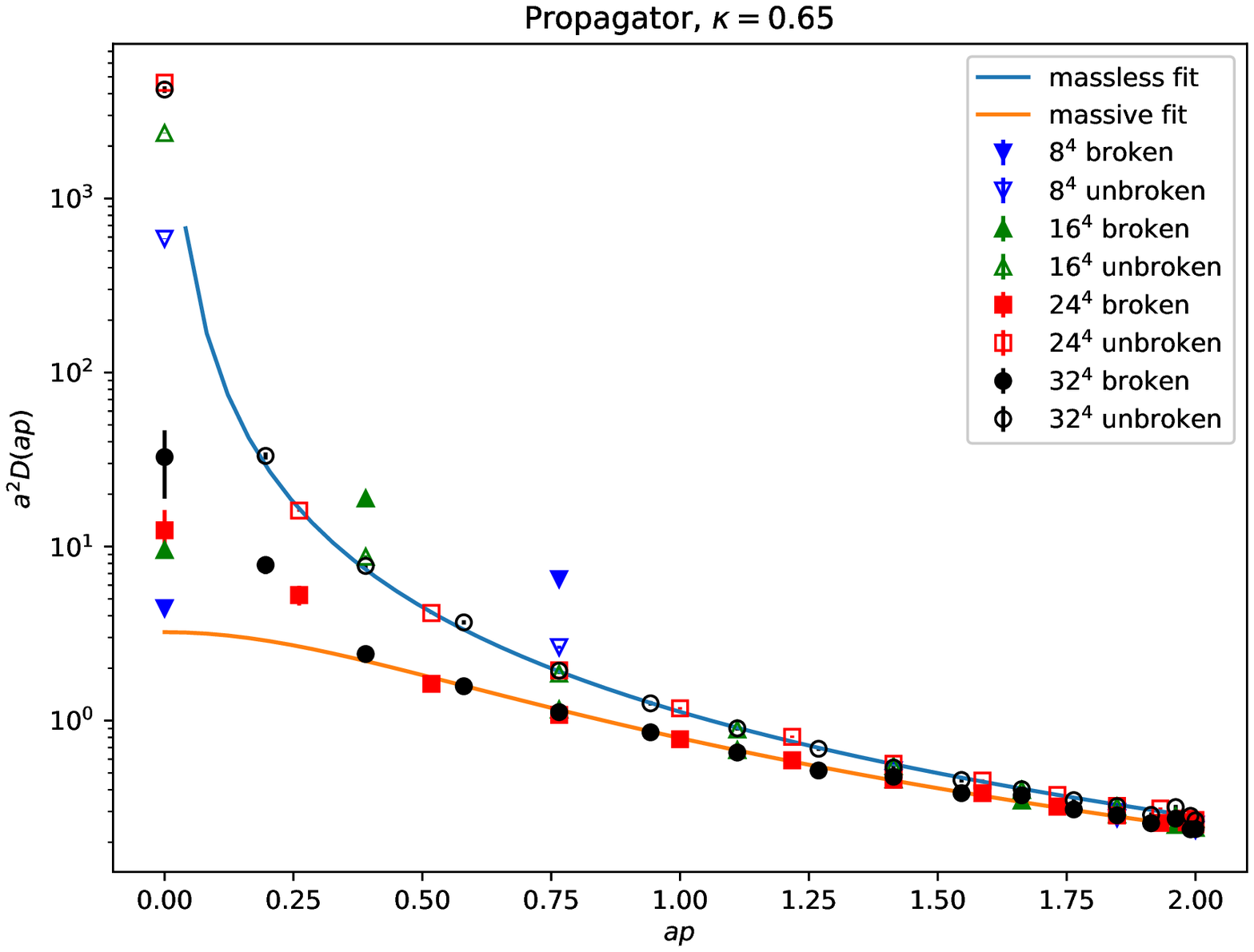}\includegraphics[width=0.5\textwidth]{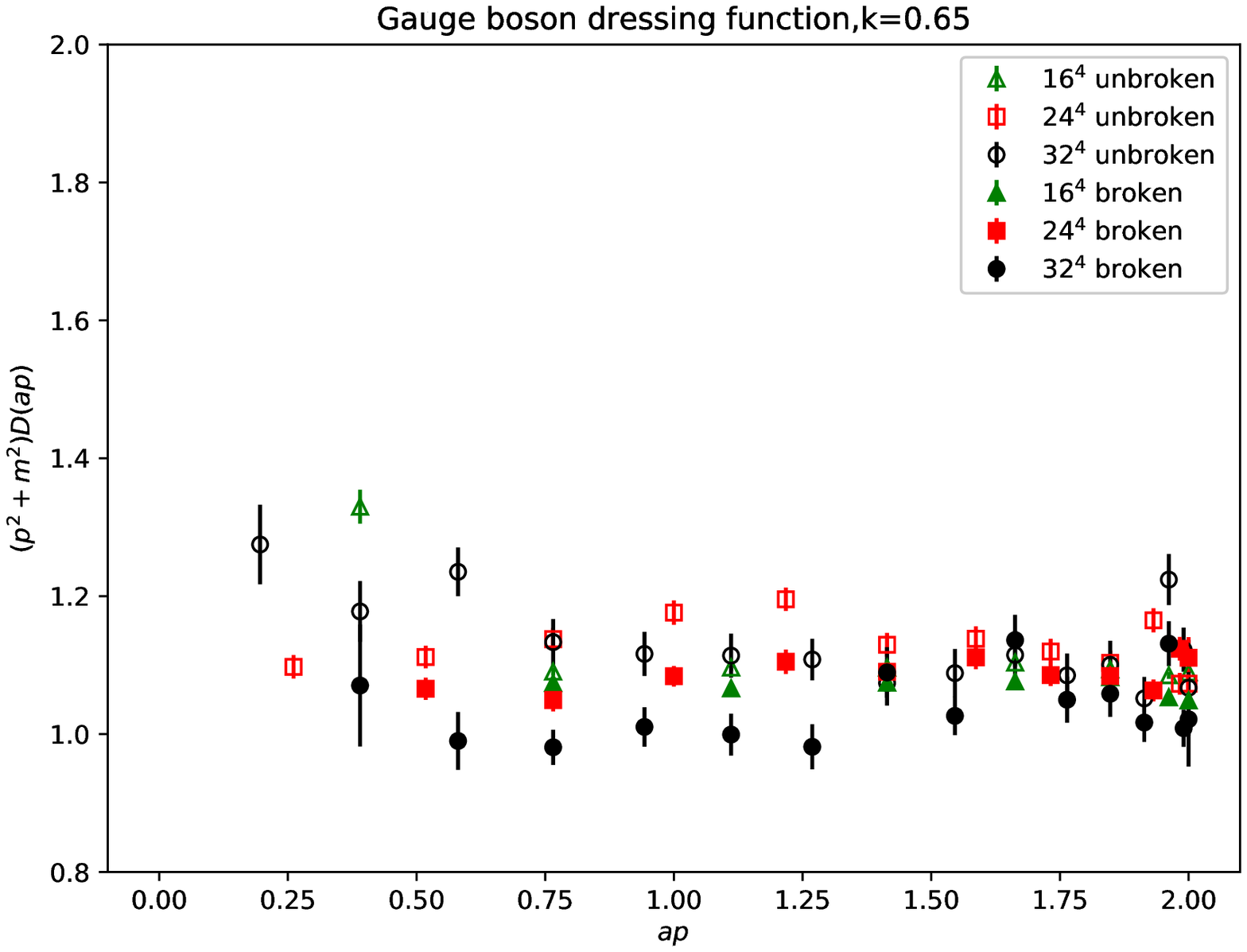}
	\caption{The gauge boson propagator (left panels) and dressing function (right panels) for $\kappa=0.55$ (top panels) and $\kappa=0.65$ (bottom panels) against tree-level fits for the $16^4$ case in lattice units. Momenta are along an edge of the lattice. The masses used to calculate the dressing functions are zero for the Cartan propagator and the fitted mass $am_A$ in Table \ref{tab:fit} for the broken sector in the right panels. Momenta are along a lattice edge.}
	\label{fig:w}
\end{figure*}

\begin{table}
\center
 \begin{tabular}{c|c|c}
 \hline
 \hline
  $L/a$ & $\kappa$ & $am_A$ \cr
  \hline
  32 & 0.55 & 0.338(1) \cr
  24 & 0.55 & 0.261(1) \cr
  16 & 0.55 & 0.207(2) \cr
  \hline
  32 & 0.65 & 0.54(2) \cr
  24 & 0.65 & 0.623(3) \cr
  16 & 0.65 & 0.585(9) \cr
  \hline
  \hline
 \end{tabular}
 \caption{\label{tab:fit}The fit parameter for the fit form \pref{fitw} of the gauge-fixed gauge boson propagator for different lattice sizes $L/a$. For the $8^4$ lattice no stable fit was possible. In figure \ref{fig:w} the values for the $16^4$ lattices have been used.}
\end{table}

The results for the gauge boson propagators for both simulation points are shown in figure \ref{fig:w}. In addition tree-level fits based on Section \ref{sub:gauge_dep_obs}
\be
D(p)=\frac{Z}{(ap)^2+(am)^2}\label{fitw}\;,
\ee
\no for the propagators are shown, with $p$ the standard improved momentum $p_\mu=2\sin(2\pi n_\mu/L)$. This fit describes the data quite well, except for the two lowest momentum points. However, the comparison of different volumes show that these points are strongly affected by finite-volume effects, and can thus be dismissed from the fits. The fit values for the masses of the massive propagator are an important ingredient in Section \ref{sec:results} and we list them therefore in Table \ref{tab:fit}. This yields that the gauge boson in the unbroken sector is indeed compatible with a massless particle, while the ones in the broken sector are compatible with tree-level massive ones. However, we observe qualitatively different, and strong, volume-dependencies for the different $\kappa$ values. This is actually consistent with the predictions from the FMS mechanism and the fact that the physical states cross various decay thresholds as a function of volume, as will be discussed in detail in Section \ref{sec:results}, and can be seen in figure \ref{fig:volume_plots}.

\begin{figure*}
	\centering
	\includegraphics[width=0.5\textwidth]{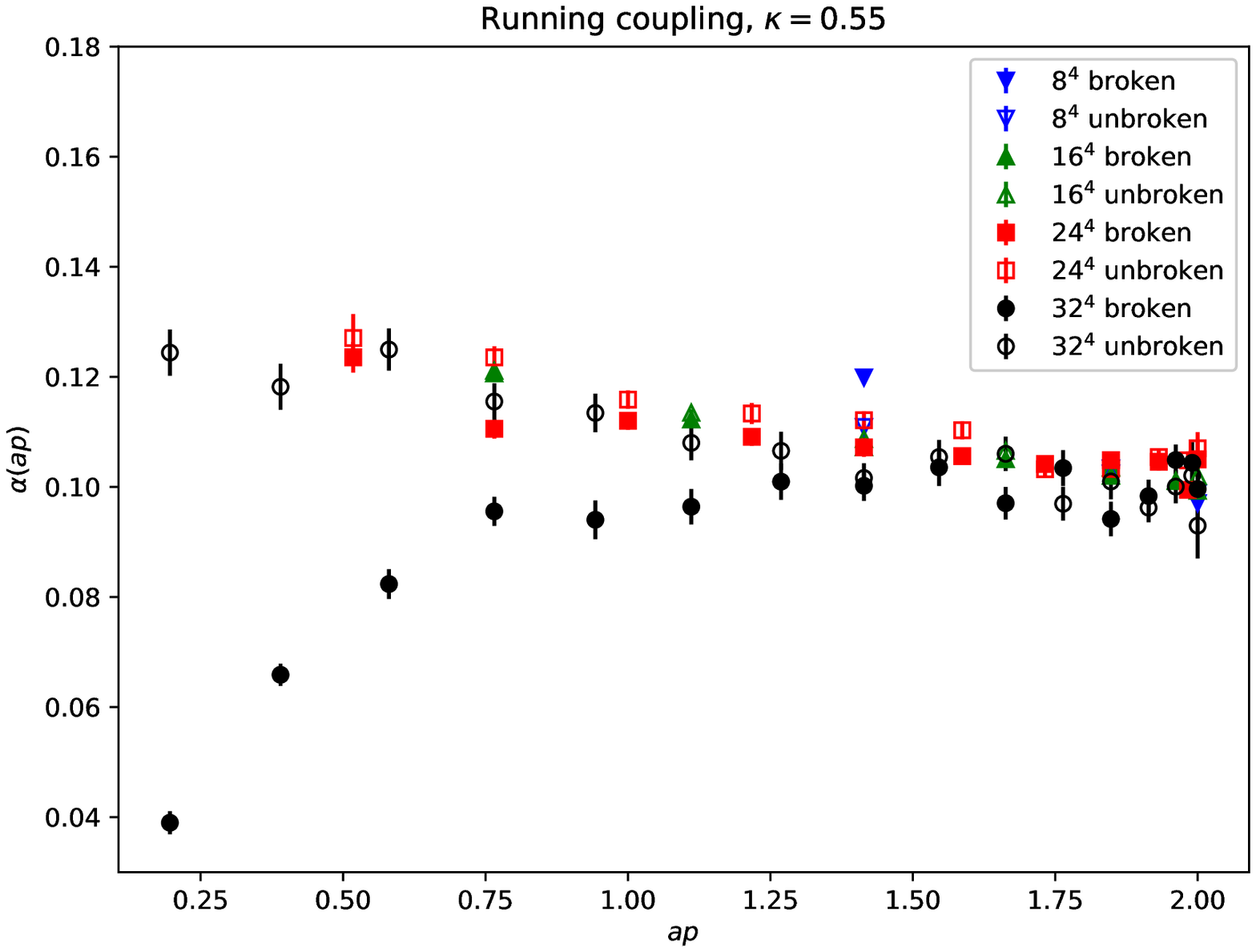}\includegraphics[width=0.5\textwidth]{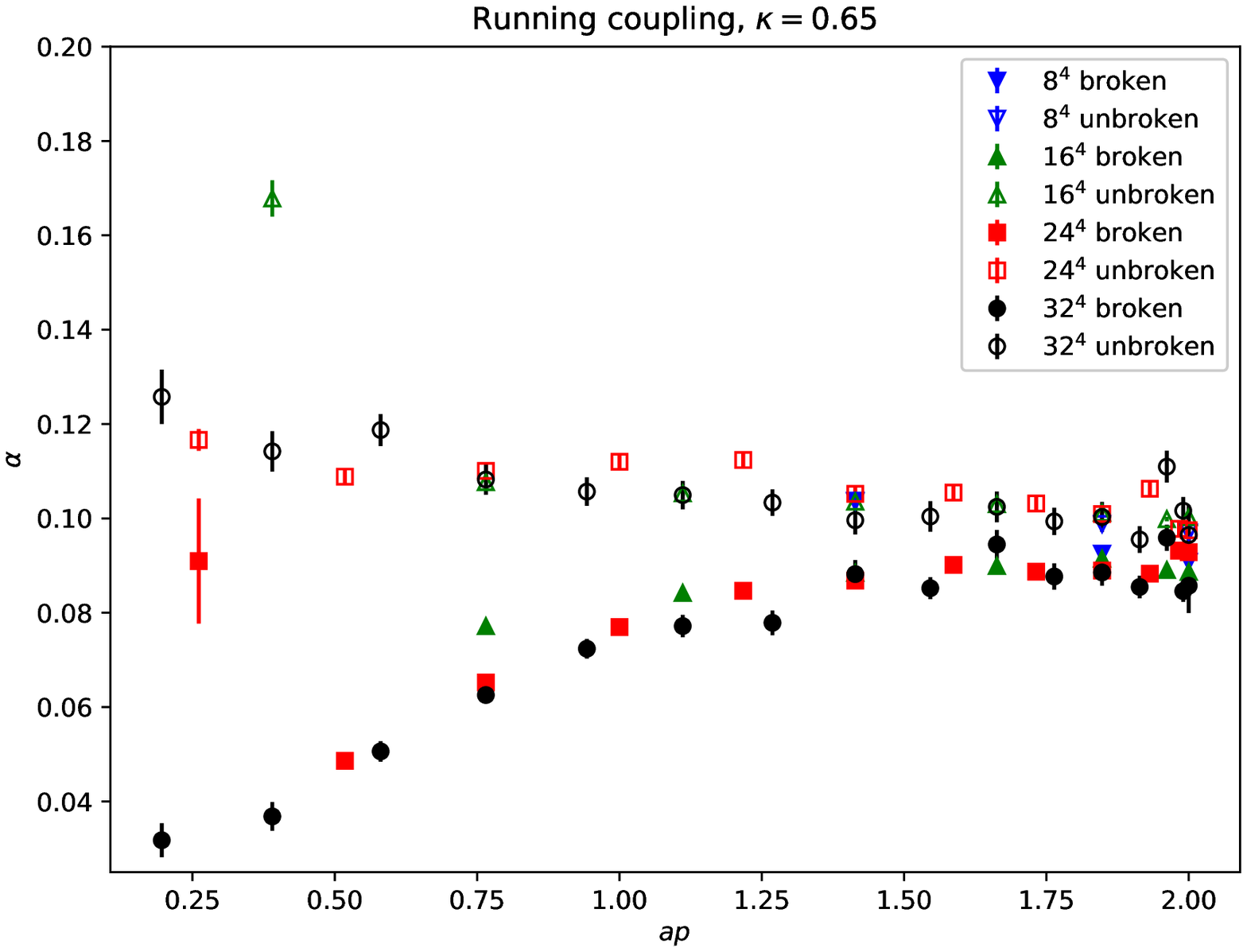}
	\caption{The running gauge coupling in the miniMOM scheme. The left panel shows the result for $\kappa=0.55$ and the right panel for $\kappa=0.65$. Note that, the lowest momentum point is very strongly affected by finite volume effects, and thus often outside the plotting range. Momenta are along an edge of the lattice.}
	\label{fig:alpha}
\end{figure*}

The running gauge coupling in the miniMOM scheme is shown in figure \ref{fig:alpha}. The picture is quite similar to the case with a fundamental Higgs \cite{Maas:2018xxu}. At large momenta the running coupling of the broken sector and the unbroken sector unifies. The momenta where they split depends on the lattice parameter, and is larger the larger $m_A$, as expected. For the lower scale with its larger volume and lower maximal physical momenta this is at $ap_\text{split}\approx 1.6$, while for the finer lattice it is at $ap_\text{split}\approx 1.1$. The lowest momenta are visibly affected by finite volume effects. Ignoring them, the coupling in the broken sector is typical for a theory with BEH effect \cite{Maas:2013aia}, and never exceeds about 0.1. For the unbroken sector, the coupling is almost momentum-independent, but also is at most $0.12$, even at the smallest momenta. Thus, both lattice settings are indeed weakly coupled, at least for the gauge interaction.

\begin{figure*}
	\centering
	\includegraphics[width=0.5\textwidth]{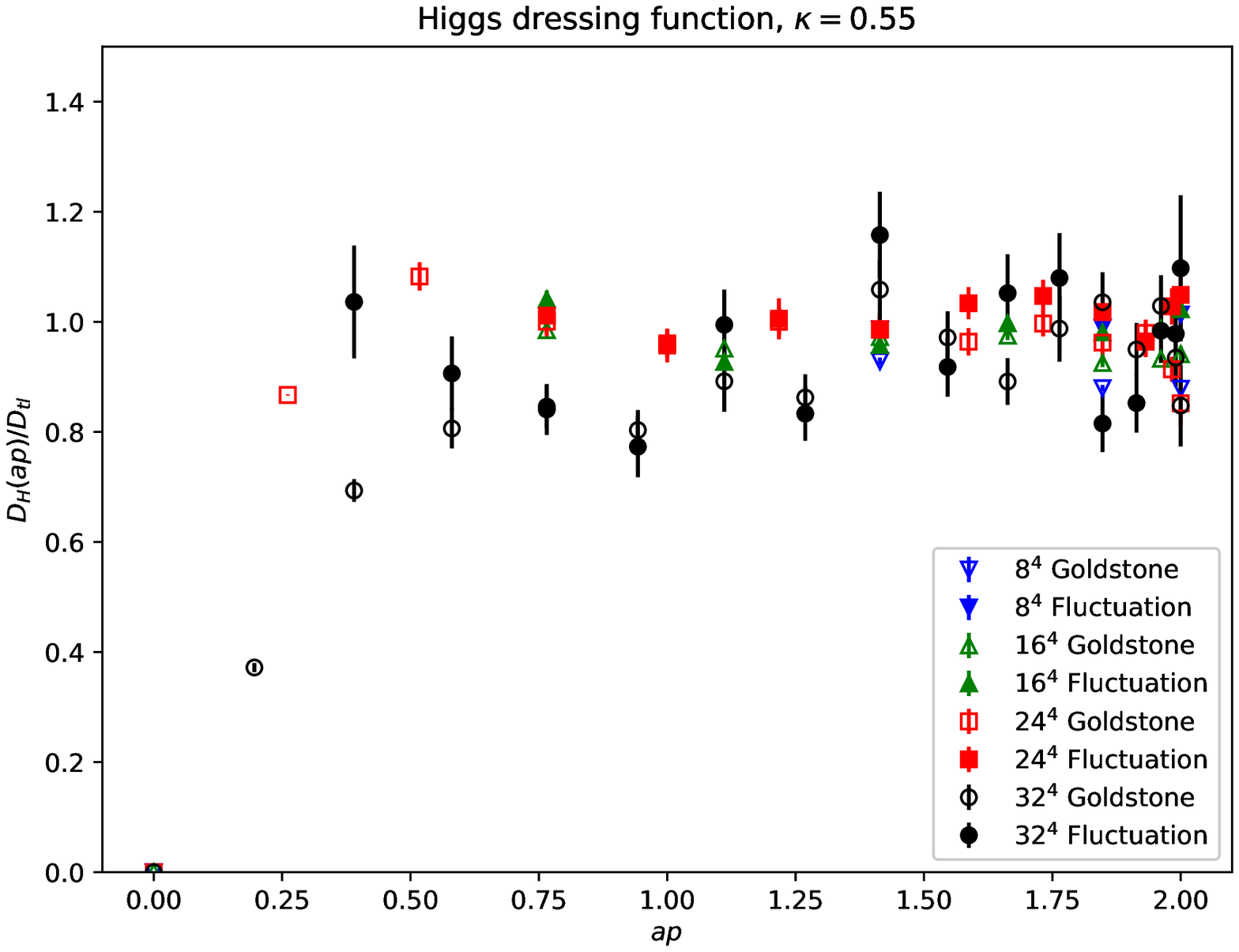}\includegraphics[width=0.5\textwidth]{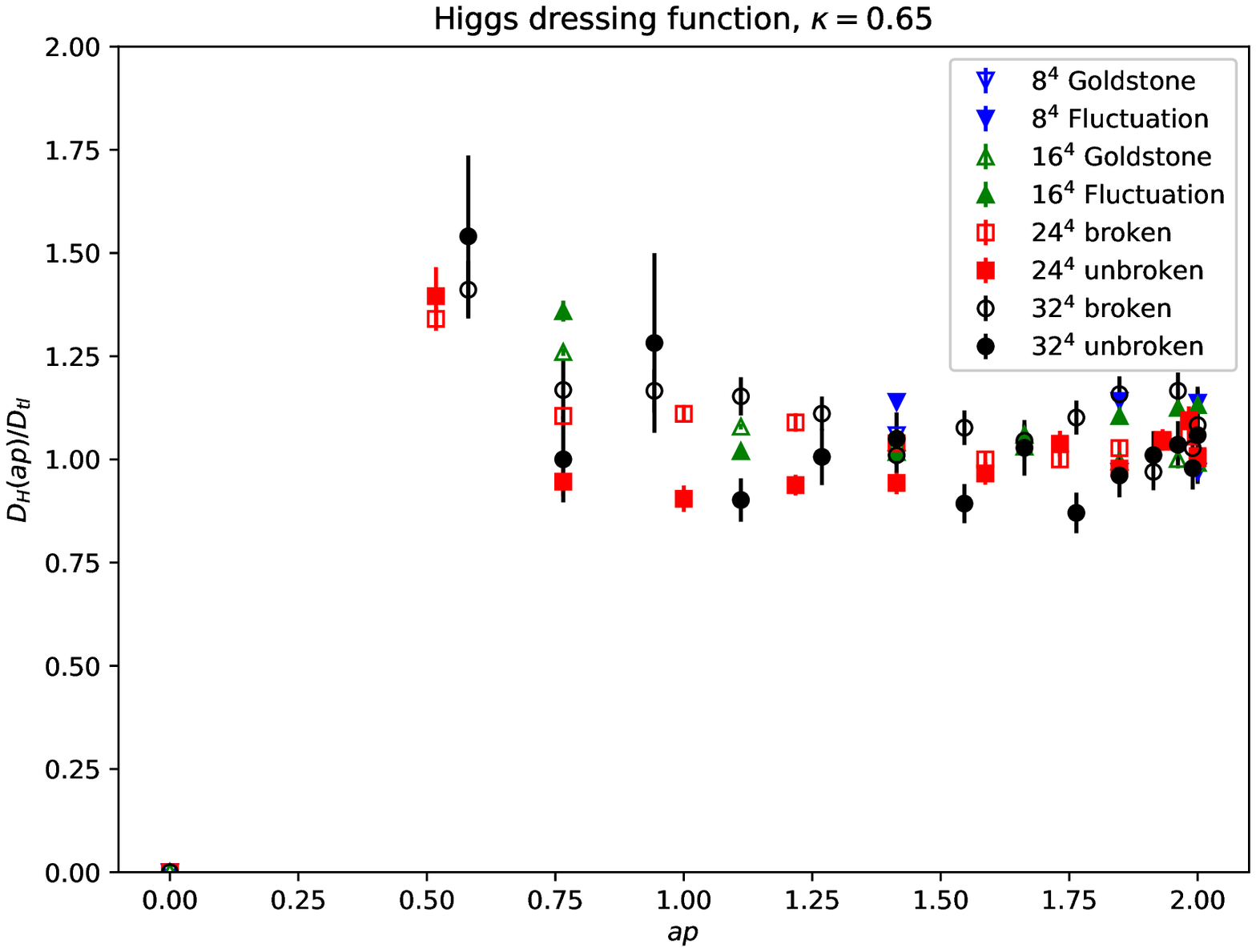}
	\caption{The renormalized Higgs dressing function normalized to the tree-level propagator in the would-be pole scheme of \cite{Maas:2010nc}. The left panel shows the result for $\kappa=0.55$ and the right panel for $\kappa=0.65$. Note that the lowest momentum point is very strongly affected by finite volume effects, and thus often outside the plotting range. Momenta are along an edge of the lattice.}
	\label{fig:h}
\end{figure*}

Finally, we show the Higgs dressing function in figure \ref{fig:h}. Because the Higgs propagator requires also a mass renormalization, we needed to choose a scheme. For this purpose, we used the one of \cite{Maas:2010nc}, which shows almost no volume-dependence \cite{Maas:2018sqz}, and appears to be a suitable approximation to the pole scheme on an Euclidean lattice \cite{Maas:2017wzi}, at least in a BEH phase. However, as we do not know the corresponding pole mass without access to the gauge-invariant scalar \cite{Maas:2017wzi}, we could only choose arbitrary masses for the renormalization condition, except for the Goldstone masses. These are massless in 't Hooft-Landau gauge. For the fluctuation propagator, we set the renormalized masses to 0.5 and 1.2 for $\kappa=0.55$ and $\kappa=0.65$, respectively. They provided reasonably stable results for all volumes, though especially the fluctuation mode on $\kappa=0.65$ turned out to be quite fickle. The resulting dressing function do not deviate substantially from the tree-level form $1/((ap)^2+(am)^2)$, and especially the Goldstone modes are well compatible with being massless. However, strong volume dependencies are also seen here at small momenta.

\section{Gauge-invariant observables}\label{sec:Gauge_invariant}

As the scalar channel is too strongly dominated by noise from the disconnected contributions, we concentrate here on the vector channel. To determine the spectrum in this channel, we employ a standard variational analysis, solving a generalized eigenvalue problem \cite{Gattringer:2010zz}. 

The following operators have been employed in this variational analysis for the study of the $J^P=1^-$ channel. All operators are averaged over time slices to reduce noise. The first operator is the simplest discretization of the continuum operator \eqref{eq:cont_operator}, see \cite{Lee:1985yi}:
\begin{equation}
\label{eq:B_operator}
B_{1^-}^i(x) = \dfrac{\mathrm{Im}\, \tr\Big[\Phi(x)\,U_{jk}(x) \Big]}{ \sqrt{2\tr\big[\Phi(x)^2\big] }}\;,
\end{equation}
where $U_{jk}$ is the usual plaquette, and the indices $(ijk)$ are even permutations of the spatial indices $(123)$. We enlarge the basis by adding two more operators
\begin{align}
B^{\Phi, i}_{1^-}(x) &= 2\,\tr\big[\Phi(x)^2\big] B_{1^-}^i(x) \;, \\
B^{2, i}_{1^-}(x) &= \Bigg(\sum_{j=1}^3 B_{1^-}^j(x) B_{1^-}^j(x) \Bigg) B_{1^-}^i(x) \;.
\label{giop2}
\end{align}
\no These represent scattering states in this channel. The first one has an insertion of another operator with quantum numbers $0^+$ constructed from the scalar field. The second one also has an insertion of a $0^+$ operator, but this one has been constructed using a product of the vector operator. Both insertions are multiplied with the operator described in \eqref{eq:B_operator} to provide the spin-parity. The additional two operators therefore describe a scattering state of a scalar and a vector, and of three vectors, respectively, with zero relative momenta.

In addition, we performed APE smearing, like in the fundamental case \cite{DeGrand:1997ss,Philipsen:1996af}, up to $n=5$ levels. The smearing procedure for the fields reads as follows:
\begin{strip}
\bea
	U^{(n)}_\mu(x) &=& \dfrac{1}{\sqrt{\det R_\mu^{(n)}(x) }} R_\mu^{(n)}(x) \;,\\
	R^{(n)}_\mu(x) &=& \alpha\,U^{(n-1)}_\mu(x) + \dfrac{1-\alpha}{6}\,\sum_{\nu \neq \mu } \Big[ U_\nu^{(n-1)}(x+ \hat{\mu})\,U^{(n-1)}_\mu(x+ \hat{\nu})^\dag\,U_\nu^{(n-1)}(x)^\dag   \nonumber \\
	&& \phantom{\,\qquad\qquad\quad}+U_\nu^{(n-1)\dag }(x+\hat{\mu} -\hat{\nu}) \,U_\mu^{(n-1)}(x-\hat{\nu})^\dag\,U_\nu^{(n-1)}(x-\hat{\nu}) \Big]\;, \nonumber \\
	\Phi^{a\,(n)}(x) &=& \dfrac{1}{7} \Bigg[ \Phi^{a\,(n-1)}(x)+ \sum_\mu \Big(V^{ab}_\mu(x)\, \Phi^{b\,(n-1)}(x+ \hat{\mu})  +  V^{ba}_\mu(x- \hat{\mu})\,\Phi^{b\,(n-1)}(x - \hat{\mu}) \Big)\Bigg] \nonumber\;,
\eea
\end{strip}
\no where $U^{(0)}=R^{(0)}$ and $\Phi^{(0)}$ describe the unsmeared fields. 
We select the tuning parameter $\alpha=0.55$, as in the fundamental case \cite{Maas:2014pba}. This created in total a maximal basis of four operators per smearing level, and 24 in total. From these we chose a subset of up to six operators, which provided for every lattice setting the least noisy results for the lowest energy levels.

\begin{figure*}
	\begin{center}
		\includegraphics[width=0.5\textwidth]{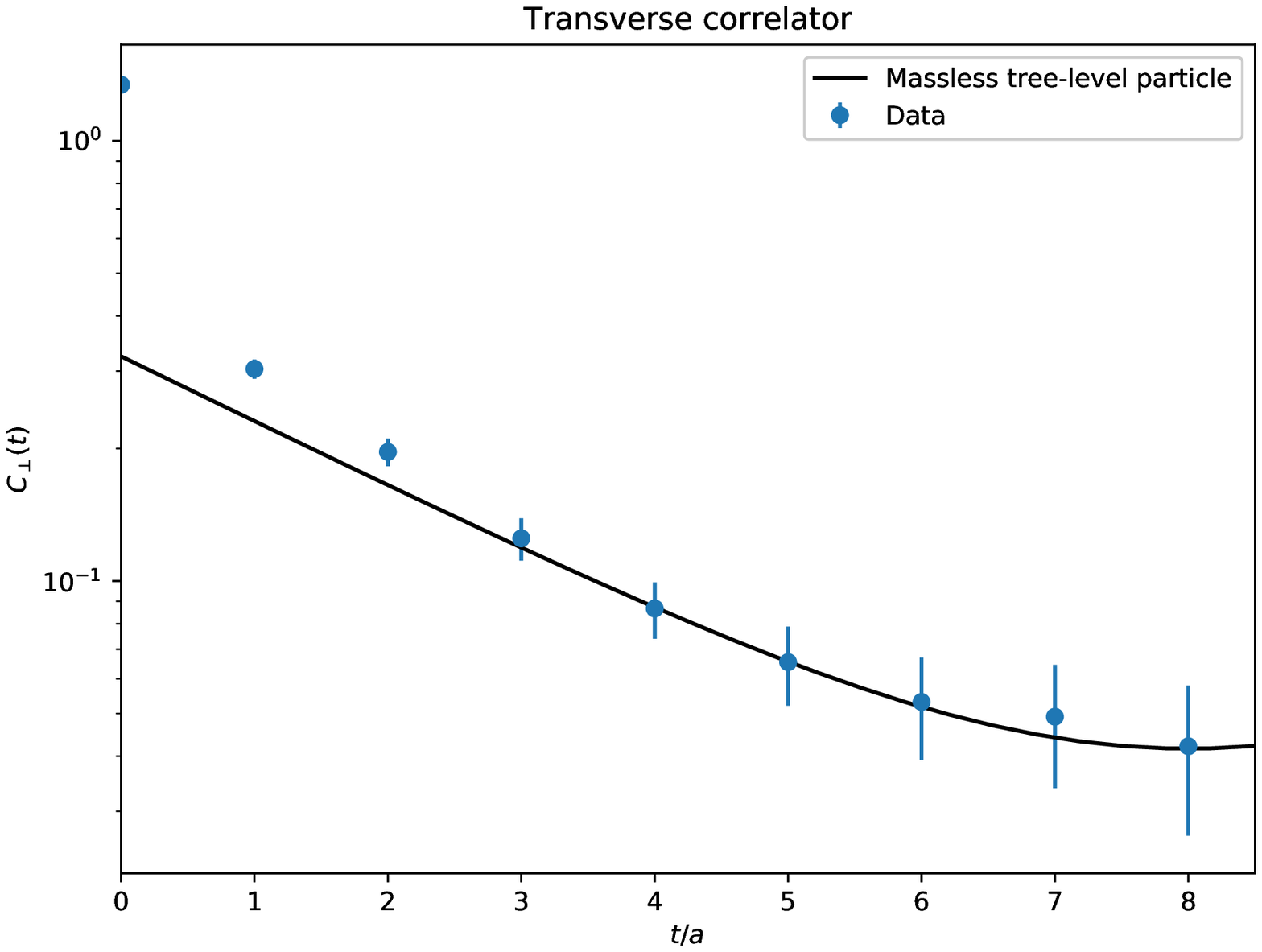}\includegraphics[width=0.5\textwidth]{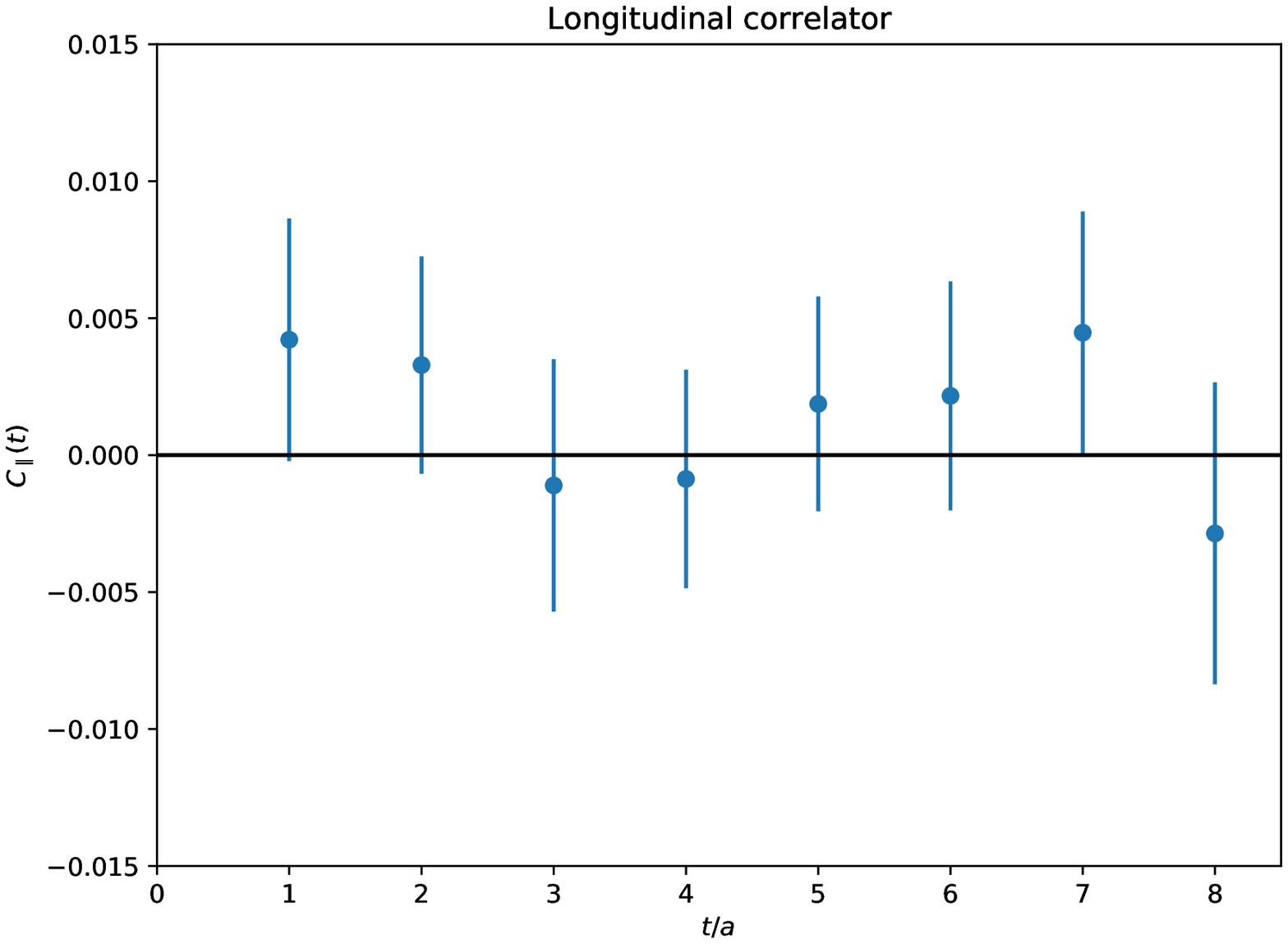}\\
		\caption{Examples for the correlator decomposition \pref{corrdecomp}, showing the transverse part (left panel) and longitudinal part (right panel) of the gauge-invariant vector correlator \eqref{eq:B_operator} in a boosted frame on a $16^4$ lattice. The simulation has been performed at $\kappa=3/4$. We also indicate the expected behavior for a massless vector particle (solid lines).}
		\label{fig:B_parallel}
	\end{center}
\end{figure*}

One particular problem is that, even in Euclidean space-time and on a finite lattice, massless vector particles cannot have a finite mass. Otherwise, a third degree of freedom would be necessary. This additional degree of freedom cannot be provided by the finite volume. Thus, to study a massless vector particle requires to work in a boosted frame\footnote{An alternative may be to use twisted boundary conditions \cite{Bornyakov:1993yy}. Note in this context also \cite{Lewis:2018srt,Lee:1985yi}.}.

Thus we boosted our operators to a non-zero momentum via 
\begin{equation}
O^j(\vec{p},t) =  \dfrac{1}{\sqrt{L^3}} \mathrm{Re}\sum_{\vec x} O^j(\vec x,t) \,e^{i \vec{p} \cdot \vec{x}} \;,
\end{equation}
with the operators $O^j$ being \eqref{eq:B_operator}-\eqref{giop2}, and it is found that also the boosted operators remain real. We chose the momentum in $z$-direction
\begin{equation}
\vec{p} = \left(0,0,p_z=\dfrac{2 \pi}{L}n_z\right)\nn\;,
\end{equation}
\no and consider $n_z=1$ for all operators. In addition, we enlarge the operator basis further by using the operator \pref{eq:B_operator} also with $n_z=2$. This turned out to be necessary to capture all relevant trivial scattering states for the analysis in Section \ref{sec:results}.

The correlators are divided in a transverse component $C_\perp$ and a longitudinal component $C_\|$ defined as
\begin{align}
C_\perp(t) &= \dfrac{1}{L}  \sum_{t'=0}^{L-1} \sum_{j=1}^2 \big\langle O^j(\vec{p}_z,t') \,O^{j}(\vec{p}_z,t +t')^\dag \big\rangle\;, \\
C_\|(t ) &= \dfrac{1}{L}  \sum_{t'=0}^{L-1}  \big\langle O^3(\vec{p}_z,t')\,O^{3}(\vec{p}_z,t +t')^\dag \big\rangle \;\label{corrdecomp}\;,
\end{align}
where time-slice averaging is performed over the points in the 4-direction of the lattice. 
 We find that the longitudinal component is zero for the ground state within statistical uncertainties, see also \cite{Afferrante:2019vsr}. This is shown for an example in figure \ref{fig:B_parallel}, and required for a massless vector particle. Hence, this is already a strong hint for the existence of a massless state in this channel. As for massive states the longitudinal component can be at most constant, we will concentrate in the following on the transverse part only.

Because we work with boosted states, we need to take the kinetic energy into account when searching for the energy levels. For this, we employ the lattice dispersion relation \cite{Gattringer:2010zz}
\begin{equation}
\label{eq:energy_lattice}
	\cosh(aE) = \cosh(a m) + \sum_{i=1}^3\big(1 - \cos(a p_i)\big) 	\;.
\end{equation}
\no Especially, in the case of massless states with a non zero momentum component $p_z$ only in the third direction the behavior should be
\begin{equation}
\cosh(a E) = 2 - \cos(a p_z) \,.\label{eq:massless_energy}
\end{equation}
\no In addition, there can be massless states with higher momenta. Furthermore, because of the perturbative and FMS predictions, we also test for other energy levels with once or twice the mass of the elementary gauge boson. In this case we can use equation \eqref{eq:energy_lattice} and the results in Table \ref{tab:fit} for the lattice energy prediction with either $m=m_A$ or $m = 2m_A$. 

\begin{figure*}
	\begin{center}
		\includegraphics[width=0.5\textwidth]{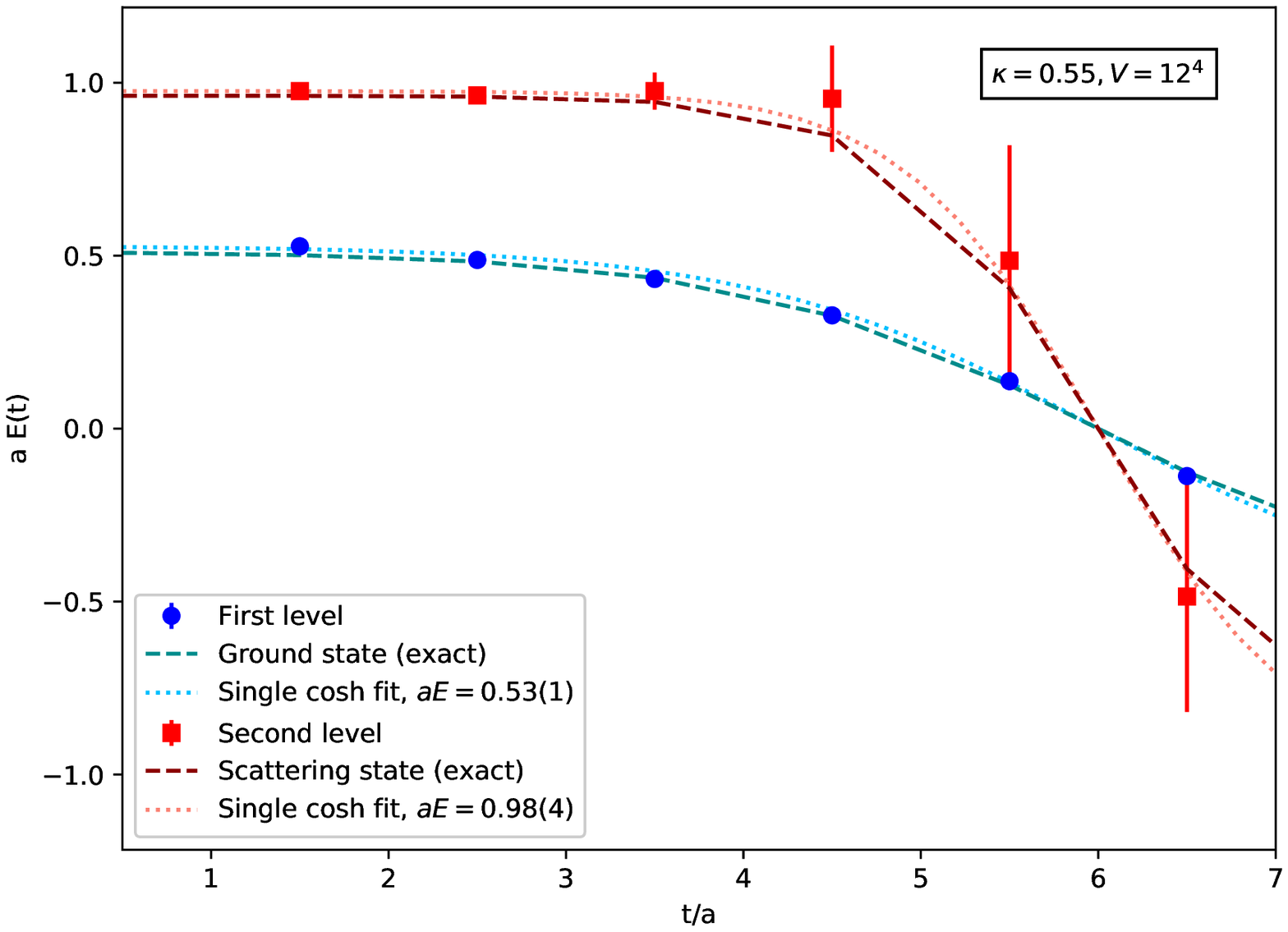}\includegraphics[width=0.5\textwidth]{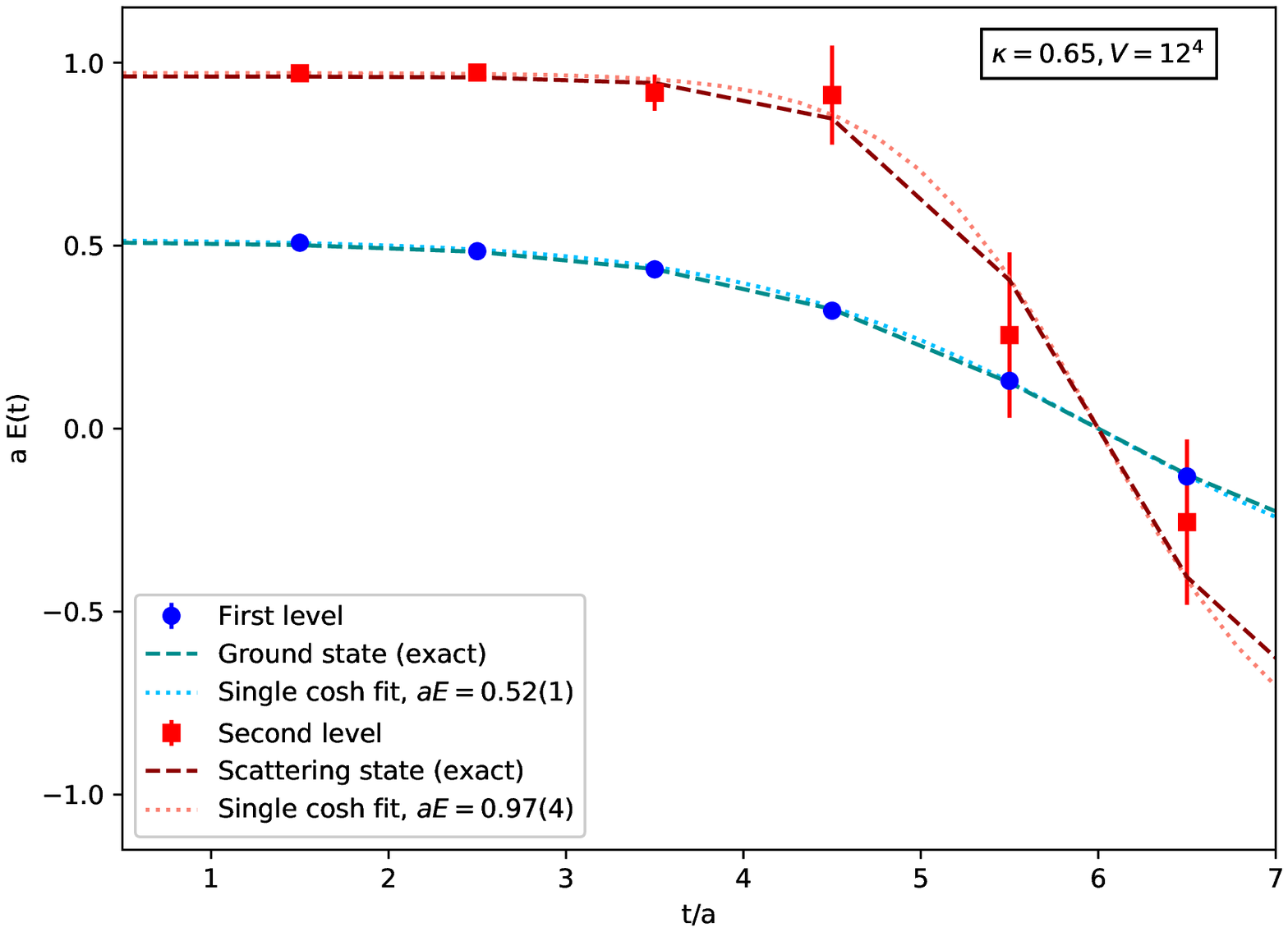}
		\caption{The plots show the effective energy obtained at $\kappa=0.55$ (left panel) and $\kappa=0.65$ (right panel) and a volume of $12^4$. They have been obtained in a basis with four operators smeared five times. Besides single-cosh fits to the data (dotted lines) also the expected behavior for a massless particle \pref{eq:massless_energy} with one unit of kinetic energy is shown (dashed lines).}
		\label{fig:energies}
	\end{center}
\end{figure*}

We demonstrate the resulting fits in figure \ref{fig:energies} for a particular lattice setup. Shown are the effective masses from the lowest eigenvalues of the variational analysis. They are compared to the expected lowest levels for a massless particle. While in this case only a single $\cosh$ was necessary for the fits, sometimes at short times the fits deviate from the expected levels due to contamination from higher levels. In these cases we included a second $\cosh$ in our fits. The resulting fits then agree very well with the expected levels at large times. Thus, our operator basis is not sufficient to disentangle very heavy states, but is suitable to identify the lowest levels quite well. Higher eigenvalues turned out to be too noisy on all but the smallest volumes, and thus we could usually only identify two levels for each volume.

\section{Spectroscopic results}\label{sec:results}

\begin{figure*}
	\begin{center}
		\includegraphics[width=0.5\textwidth]{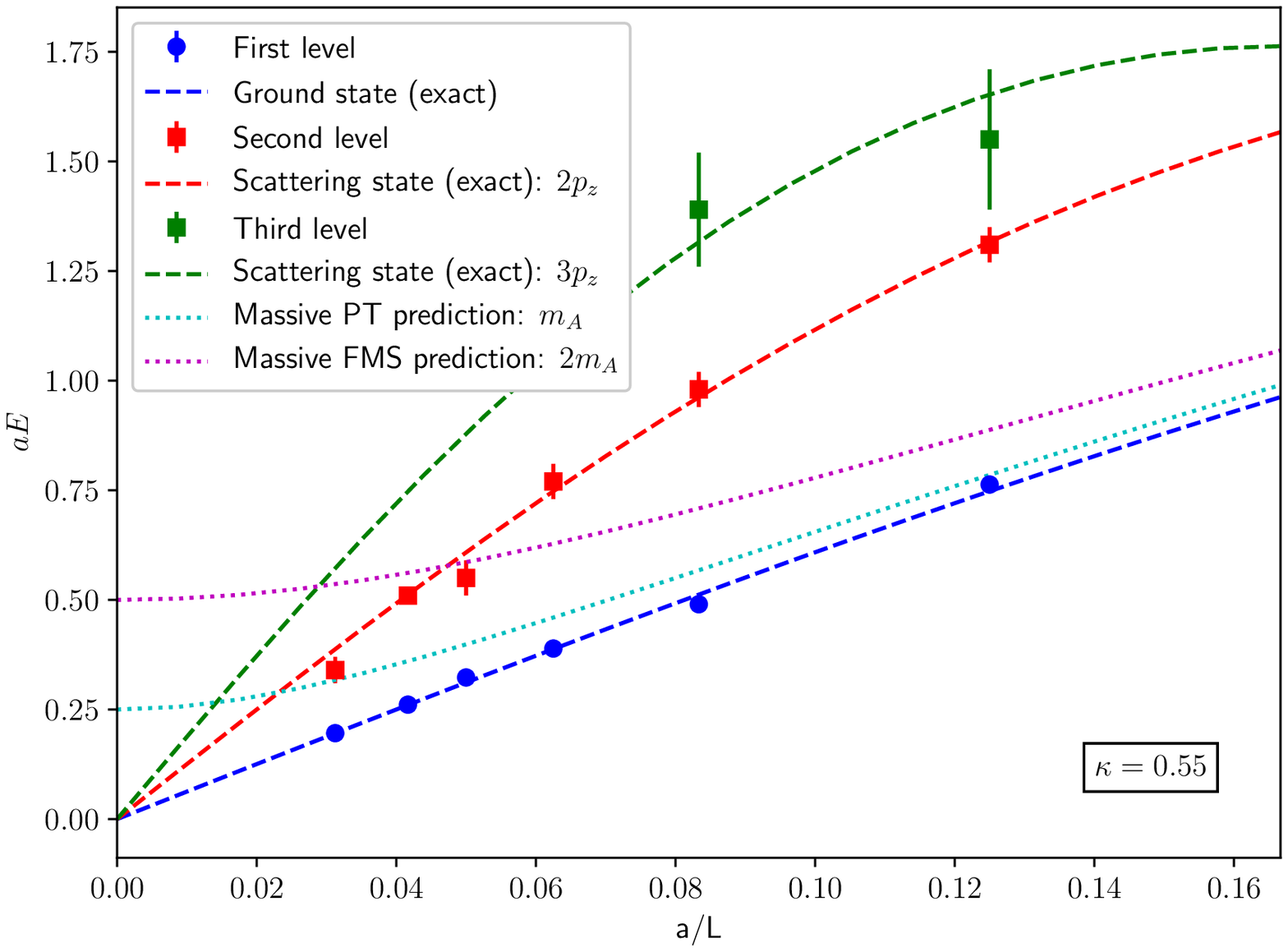}\includegraphics[width=0.5\textwidth]{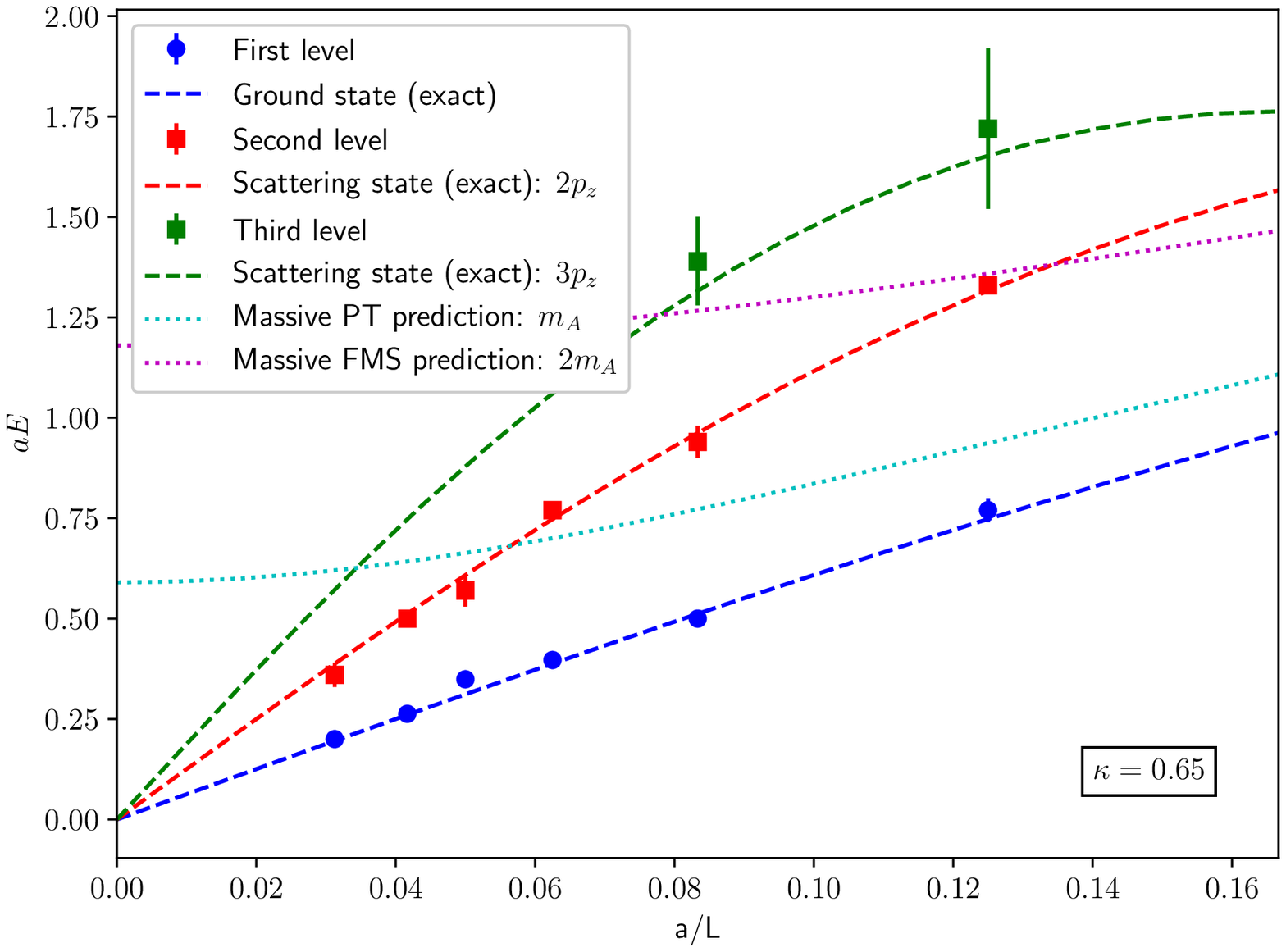}
		\caption{\label{fig:volume_plots}The plots show the volume-dependent low-lying spectrum for $\kappa=0.55$ (left) and $\kappa=0.65$ (right). Besides the simulation results also shown are the predictions for the three lowest-lying massless states (dashed lines), the massive state from perturbation theory (cyan dotted lines) and the FMS prediction for an additional massive state (magenta dotted lines). In the latter cases masses $am_A=0.25$ and $am_A=0.6$, respectively, have been used, as reasonable proxies to the masses in Table \ref{tab:fit}. For the massive predictions effects from avoided level crossing have not been included in this plot.}
	\end{center}
\end{figure*}

Before studying the final results, it is worthwhile to list the expectations. On the one hand, there should be a massless state. In our boosted frame we expect it to have energies corresponding to one or more units of kinetic energy, which behave like $aE_{n_z}\approx 2\pi n_z/L$. In addition, there are two different predictions for massive states. The one from perturbation theory should have a mass $m_A$, while the one from the FMS mechanism should have $2m_A$. In the boosted frame, both will have at least one unit of kinetic 
energy $E_1$ as well. In addition, any massive state of mass $am$ in this frame can only decay into at least three massless ones. Thus, this is only possible if
\be
3+\cosh(a m)-\cos\left(\frac{2\pi}{L}\right)<2-\cos\left(\frac{6\pi}{L}\right)\nn
\ee
\no is satisfied. Though the masses show some volume-dependence, this effect is dominated for our lattice setups by the volume-dependence of the kinetic energy. As a function of volume, both predicted massive states eventually cross the elastic decay threshold when increasing the volume, though at different ones.

Note that adding the perturbative state directly does not make full sense, as it has different quantum numbers: It is charged under the residual gauge U$(1)$. Thus, it cannot be observable at all. However, it could be argued that it should still be manifest in the spectrum, by dominating some other state. Its absence is again a prediction of the FMS mechanism \cite{Maas:2017xzh}, which warrants checking.

The final results are shown in figure \ref{fig:volume_plots}, compared to these expectations. While we were not able to extract more than the two lowest-lying states on all volumes, we see a rather clear picture emerging.

First, the ground state is throughout consistent with the expected massless state. Hence, the ground state in the vector channel in this theory is pretty likely a massless, composite particle. Thus, this basic prediction of a composite massless vector from the FMS mechanism is confirmed. We also see very clearly and consistently a state which is compatible with a massless state with two units of kinetic energy. Thus, the existence of a massless, composite vector particle in this theory is well supported.

We do, however, not see any indications of either of the massive states. Especially, we do not see any hints of these states even on volumes were we would expect them to be stable, as they are below the corresponding decay threshold, the third massless level, which is also indicated in figure \ref{fig:volume_plots}. We also see no deformation indicative of avoided level crossing or additional states. Thus, at the moment, neither of the additional massive states is seen.

The reason for this may, of course, be the operator basis, which always included the primitive operator \eqref{eq:B_operator}. Other operators \cite{Sondenheimer:2019idq} may be needed, e.g., like those employed in \cite{Wurtz:2013ova}. Unfortunately, for massless (vector) particles no L\"uscher analysis is (yet) available to check for possible resonances. There is, of course, also the possibility of further discretization artifacts, finite volume effects, or too little statistics for only small admixtures. Such improvements would be straightforward, but would require substantially more computing time.

If even such extensions would not detect theses states, this would have different implications. In perturbation theory, the (unstable) massive vector state is unambiguously predicted. Its absence would therefore be in direct contradiction to perturbation theory. In the FMS approach, this would invalidate the simplified constituent model in \cite{Maas:2017xzh}, but may be understood in a more advanced analysis \cite{Sondenheimer:2019idq} yet to be performed.

\section{Conclusions}\label{sec:conclusion} 

Summarizing, we have obtained substantial evidence for a massless, composite vector state in the Brout-Englert-Higgs regime of the SU(2) theory with a Higgs in the adjoint representation. This confirms the exploratory study \cite{Lee:1985yi}. Moreover, we find no indications for additional massive states. The latter would, however, necessarily be resonances in the infinite-volume limit. 

We have thus provided evidence that such a theory can create, in a manifestly gauge-invariant way, a particle which could be regarded as a low-energy effective photon in a grand-unified-theory setting. This is needed to obtain a non-perturbatively gauge-invariant construction of a GUT \cite{Maas:2017xzh,Sondenheimer:2019idq}. In addition, this is also a proof-of-principle that massless non-scalar bound states can emerge without a broken (global) symmetry, and thus not as a Goldstone boson. This may be an interesting option also in other extensions of the standard model, and may also be relevant to quantum gravity \cite{Maas:2019eux}.

In addition, by comparison to the gauge-fixed vector particles, we support the analytic prediction for the bound state spectrum in the vector channel by the FMS mechanism for this theory for the ground state \cite{Maas:2017xzh}. That the ground-state comes out correctly in such calculations is by now familiar from other theories \cite{Maas:2018xxu,Maas:2013aia,Maas:2012tj,Maas:2016ngo}. However, we do not see additional massive states with non-trivial internal structure, which have been argued for \cite{Maas:2017xzh,Sondenheimer:2019idq}. Only for trivial internal structure this has so far happened,  experimentally confirmed, in the standard model for the photon and the $Z$-boson \cite{Maas:2017wzi}.

In total, these results are therefore a vital step towards a fully gauge-invariant construction of a GUT, and another example that FMS-mechanism augmented perturbation theory is the best method to deal with (non-Abelian) gauge theories involving the Brout-Englert-Higgs effect.

Nonetheless, a full determination of the spectrum in other channels remains desirable for the outlined gauge-invariant description of GUTs. A logical next step is therefore to focus on the scalar channel in the future. Understanding the scalar channel would potentially also help to shed more light on the results in the vector channel, and is a necessary input for further analytic calculations in FMS-augmented perturbation theory.

\section*{Acknowledgments}

We are grateful to C.\ B.\ Lang and R.\ Sondenheimer for useful discussions, to R.\ Sondenheimer also for a critical reading of the manuscript, and to C.\ Pica for providing us with the HiRep code in early stages of this work. V.\ A.\ is supported by the FWF doctoral school W1203-N16. The computations have been performed on the HPC clusters at the University of Graz and the Vienna Scientific Cluster (VSC).

\begin{appendices}
\section{Thermalization properties of the algorithm}\label{Apppendix:therm}

As noted in Section \ref{sub:pd}, we found that the theory is very hard to thermalize, especially when being deep in the BEH region. Given the observation of the massless mode, this does not come as a significant surprise, as light modes usually yield long correlations.

Originally, we started this project using a modified \cite{Hansen:2017mrt} variant of the HiRep code \cite{DelDebbio:2008zf}. This code is based on a hybrid Monte Carlo. We have augmented it to deal with the adjoint Higgs. For this purpose, we used various decompositions of the Higgs field. Especially we explicitly attempted to decouple the radial and the angular mode. We found that this algorithm suffered from a lack of thermalization for values of $\kappa$ larger than $0.2$. Especially, for all practical purposes even volumes as small as $24^4$ effectively no longer thermalized. The algorithm required an extensive amount of time for updates and in general a really low acceptance rate for the new proposed configurations in the regime with $\kappa$ larger than 0.2. This applied both to the vacuum expectation value of the Higgs, but even to local quantities like the plaquette. We are not sure what precisely created this behavior, but we suspect that the attempted global update in the hybrid Monte-Carlo yielded only too small steps inside the potential trough of the Higgs, and could therefore not move efficiently.

\begin{figure}
	\centering
	\includegraphics[width=0.4\textwidth]{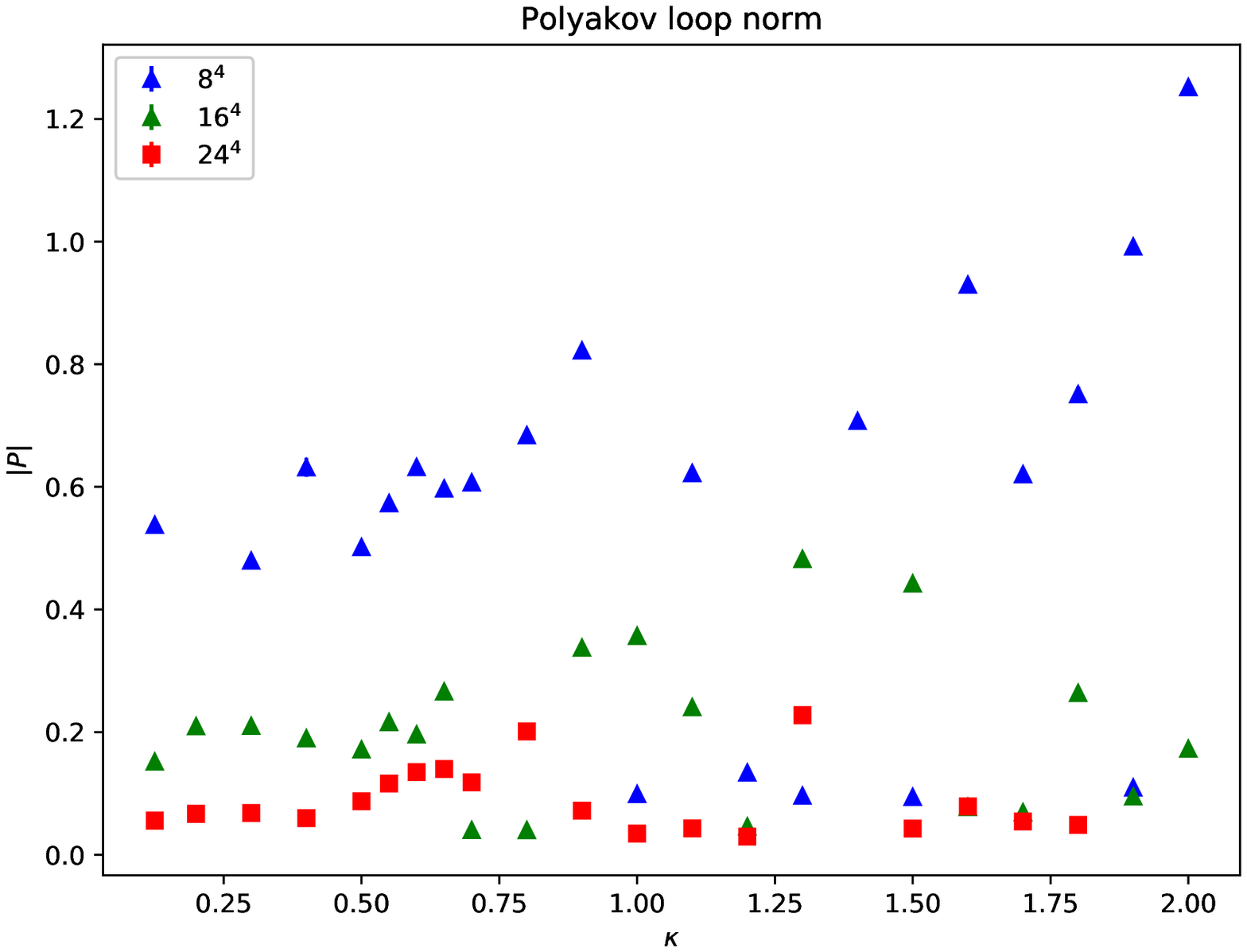}\\
	\includegraphics[width=0.4\textwidth]{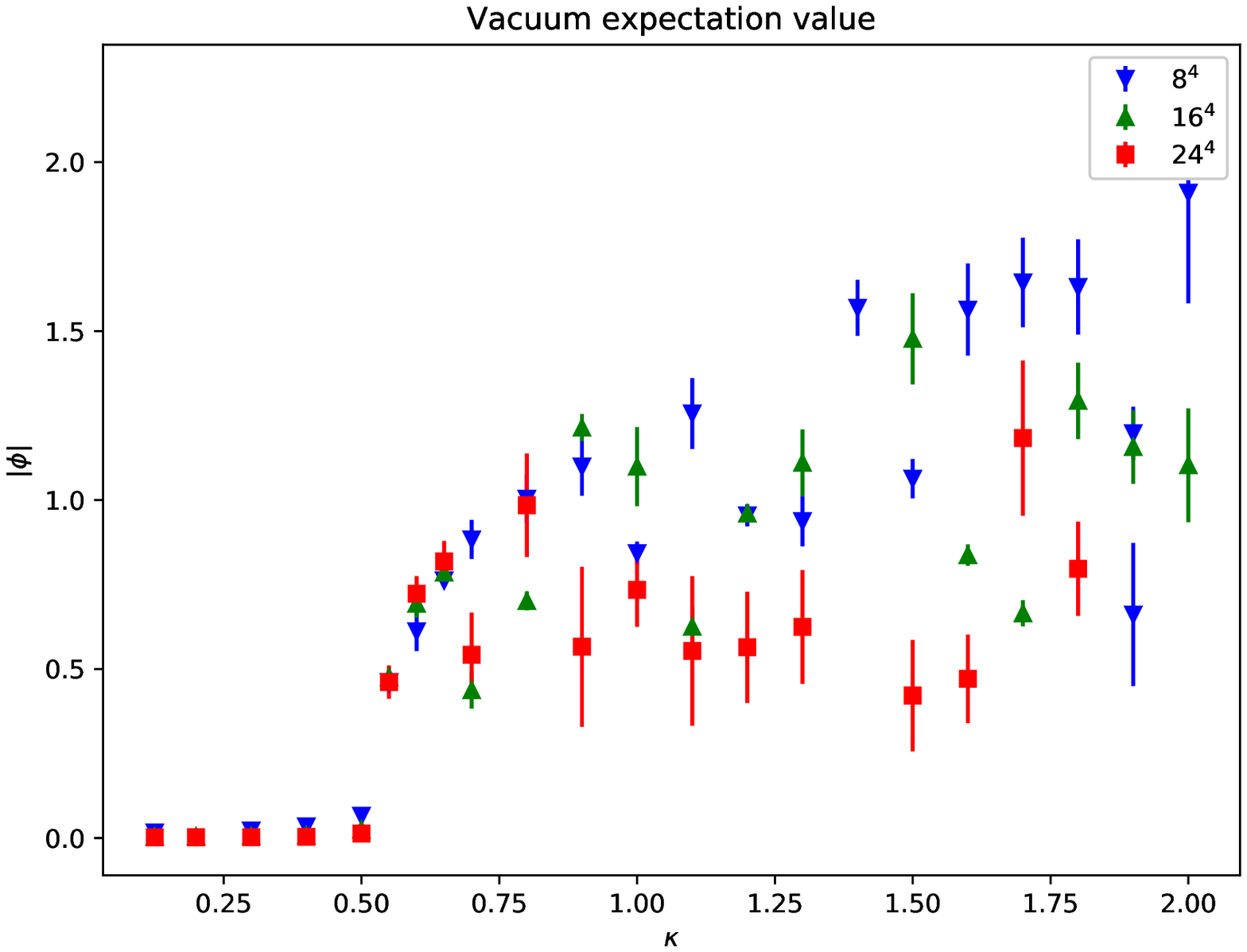}\\
	\includegraphics[width=0.4\textwidth]{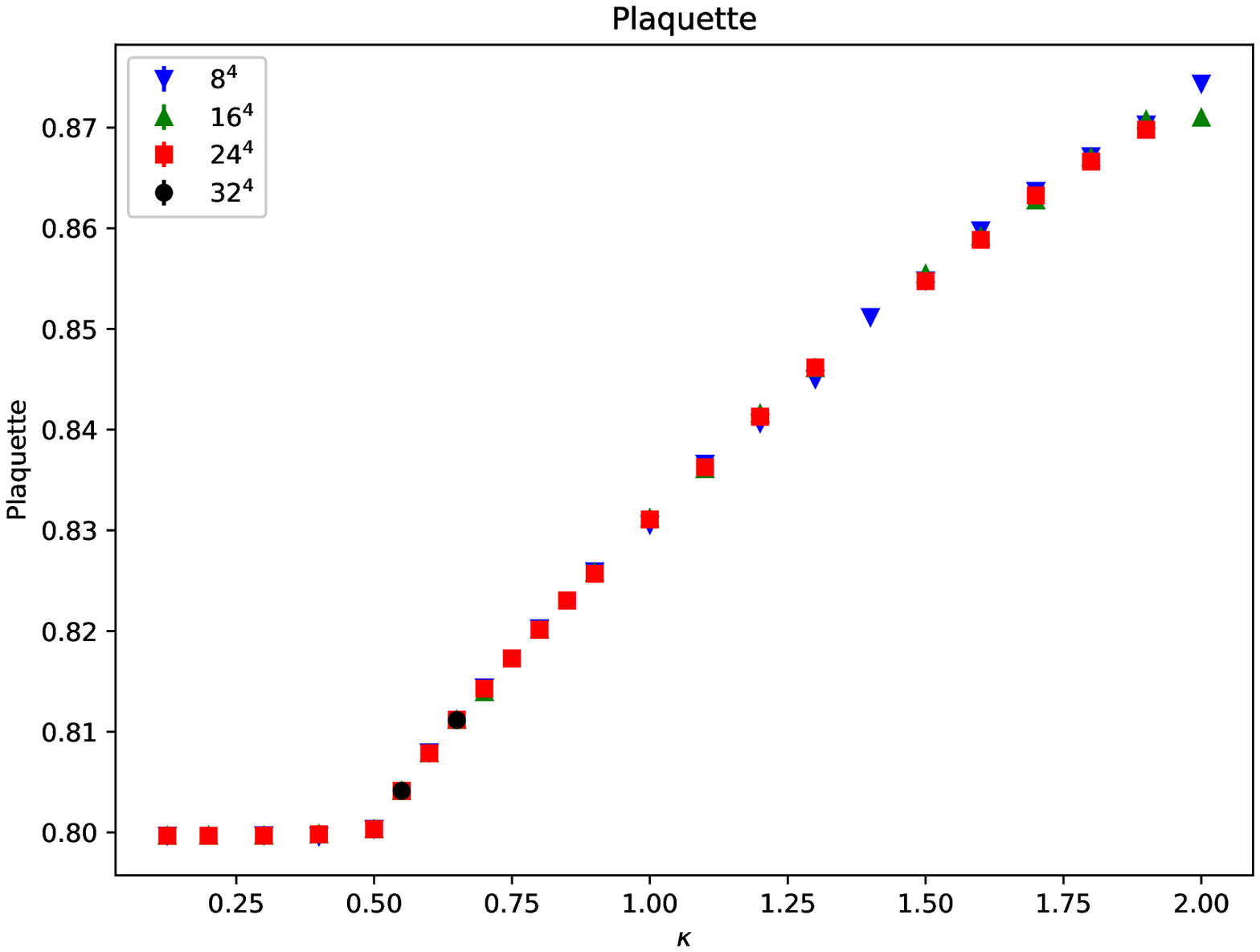}
	\caption{The norm of the Polyakov loop (top panel), the Higgs vacuum expectation value (middle panel), and the plaquette (bottom panel) for various volumes as a function of $\kappa$ for $\beta=4$ and $\lambda=1$. For the Higgs vacuum expectation value the statistical error has been enlarged by a factor of ten to demonstrate that the observed effect is definitely not a statistical problem.}
	\label{fig:therm-prob}
\end{figure}

We thus reverted to a local algorithm, a multi-hit Metropolis algorithm, as was already successfully used previously for Yang-Mills-Higgs systems \cite{Maas:2016ngo,Maas:2018xxu}. This proved successfull also in our case, allowing us to perform simlations with set of parameters which were practically inaccessbile with the previous algorithm. However, we found that even in this case thermalization became problematic at too large values of $\kappa\gtrsim 0.7$. This is shown in figure \ref{fig:therm-prob}. It is visible that the ultra-local plaquette behaves now well throughout, but both the Higgs vacuum expectation value and the Polyakov loop, which are both objects obtained from non-local quantities, are not. It must be stressed out that the natures of the thermalization issues in the two cases are quite different. In general, we observed much more severe difficulties in generating the configuration with a global update, since with increasing values of $\kappa$ the acceptance rate decreased significantly, while this effect was much less harsh with local updates.

\begin{figure}
\centering
\includegraphics[width=0.5\textwidth]{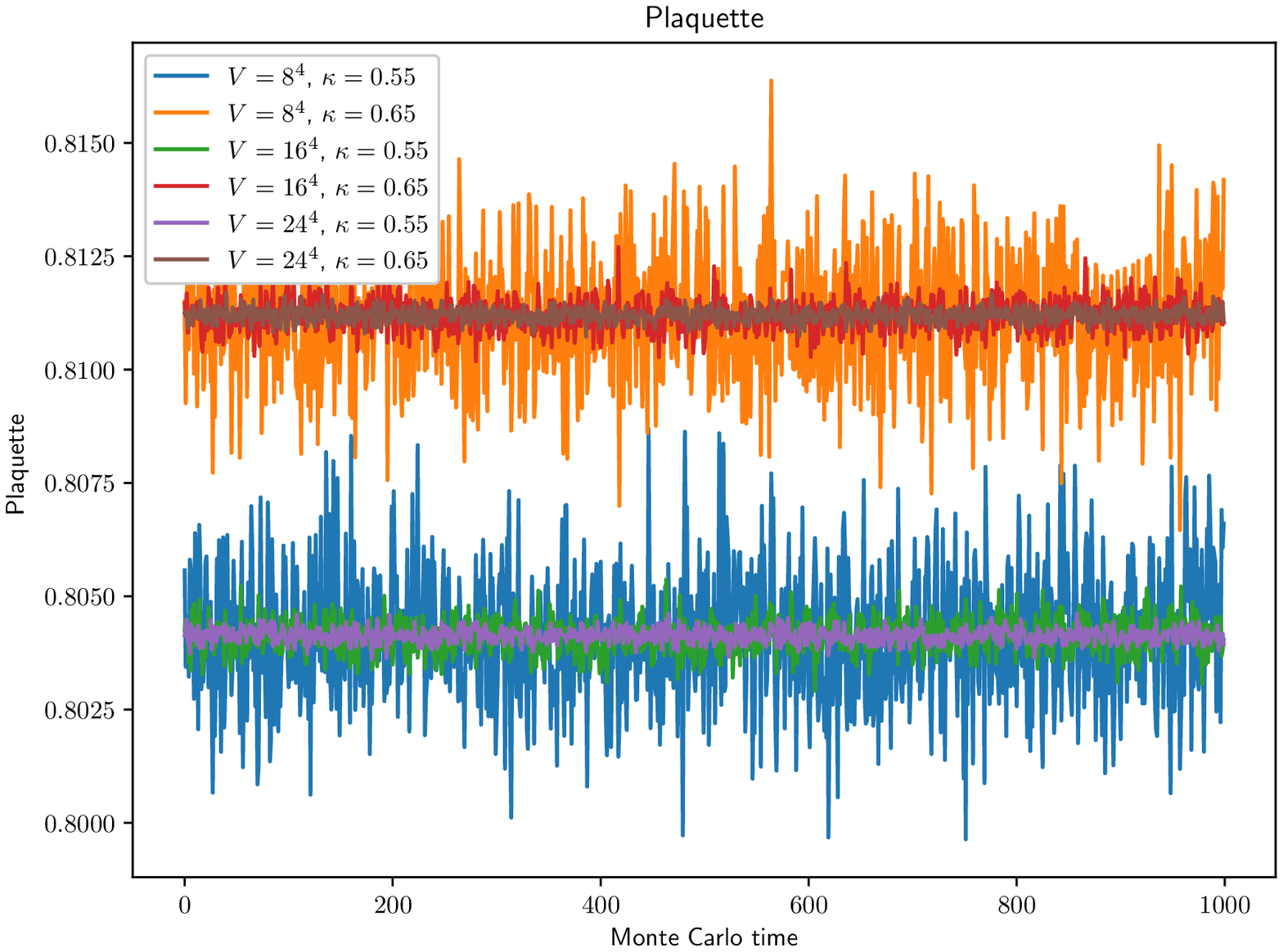}
\includegraphics[width=0.5\textwidth]{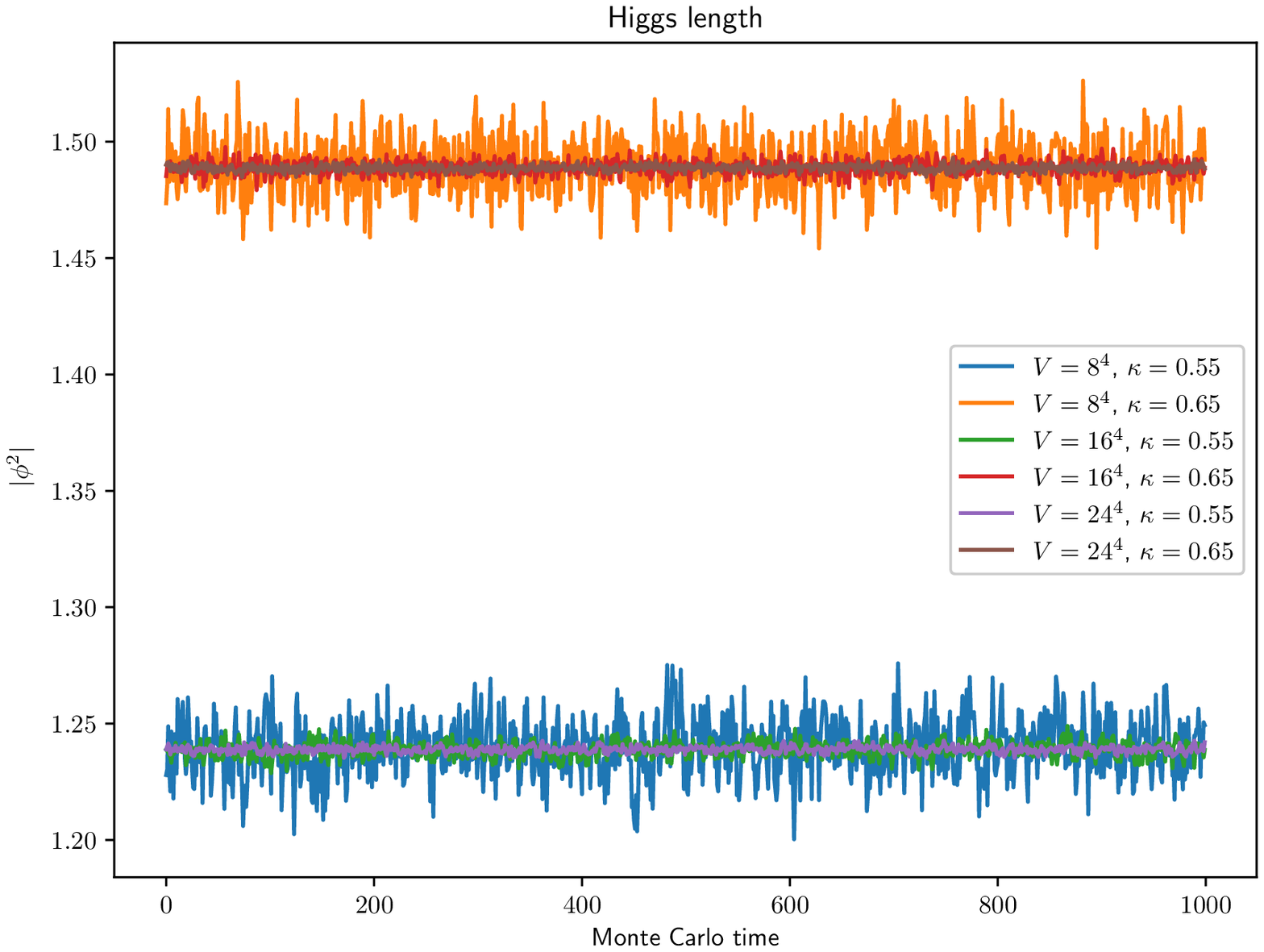}
\caption{Monte Carlo time evolution of the first 1000 measurements of the plaquette (top panel) and Higgs length (bottom panel) for various volume and $\kappa=0.55$ and $\kappa=0.65$. The measurements have been done with the same configurations used for the spectroscopical analysis in the main text.}
\label{fig:timestamps}
\end{figure}

When investigating Monte-Carlo trajectories, we find that the reason for the jumping behavior comes from excursions to configurations with vastly different values of the Higgs vacuum expectation value and the Polyakov loop norm. Occasionally, it also happens that the algorithm gets stuck. Since the plaquette seems to change discontinuously, this could be due to a two-state system. However, neither of the phases shows a vanishing Higgs vacuum expectation value, as is also visible in figure \ref{fig:therm-prob}. Also, it would usually not be expected that this becomes a stronger problem further away from the phase transition. We therefore expect that this is still a sign of slow thermalization, which allows for large excursions in configuration space. This is also consistent with the observation that the values of the observables in the various trajectories seems to be rather random. We therefore conclude that also our multi-hit Metropolis algorithm is not able to thermalize quickly enough for $\kappa\gtrsim 0.7$, and hence restrict ourselves to smaller values of $\kappa$ in the main analysis. This is exemplified in figure \ref{fig:timestamps}, showing  the Monte-Carlo evolution of the plaquette and the Higgs length. They are measured using the same configurations employed in the spectrocopical analysis, and no signs of thermalization issues are present. Especially, none of the excursions of the plaquette to different values observed at $\kappa\gtrsim 0.7$ are seen.

\end{appendices}

\bibliographystyle{bibstyle}
\bibliography{bib}


\end{document}